\documentclass{aa}
\usepackage{psfig}
\usepackage{natbib}

\def\h2{H{\small II}}

\newcounter{qub}
\begin{document}

\title{The chemical composition of metal-poor emission-line galaxies
in the Data Release 3 of the Sloan Digital Sky Survey\thanks{Tables 
\ref{tab1} and \ref{tab2} 
are only available in electronic form
at the CDS via anonymous ftp to cdsarc.u-strasbg.fr (130.79.128.5)
or via http://cdsweb.u-strasbg.fr/cgi-bin/qcat?J/A+A/}
}

\author{Y. I. Izotov \inst{1}
\and G.\ Stasi\'nska \inst{2}
\and G.\ Meynet \inst{3}
\and N. G.\ Guseva \inst{1}
%\and L. S.\ Pilyugin \inst{1}
\and T. X.\ Thuan\inst{4}}
\offprints{Y. I. Izotov, izotov@mao.kiev.ua}
\institute{      Main Astronomical Observatory,
                     Ukrainian National Academy of Sciences,
                     Zabolotnoho 27, Kyiv 03680,  Ukraine
\and
                     LUTH, Observatoire de Meudon, F-92195 Meudon Cedex, France
\and
                      Geneva Observatory, CH-1290 Sauverny, Switzerland
\and
                     Astronomy Department, University of Virginia,
                     Charlottesville, VA 22903, USA
%Georges.Meynet@obs.unige.ch
}

\date{Received \hskip 2cm; Accepted}

\abstract{We have re-evaluated empirical expressions for the abundance
determination of N, O, Ne, S, Cl, Ar and Fe taking into account the latest
atomic data and constructing an appropriate  grid of photoionization models
with state-of-the art model atmospheres. Using these expressions
we have derived heavy element abundances in the $\sim$ 310 emission-line
galaxies from the Data Release 3 of the Sloan Digital Sky Survey (SDSS)
with an observed H$\beta$ flux $F$(H$\beta$) $>$
10$^{-14}$ erg s$^{-1}$ cm$^{-2}$ and for which the [O {\sc iii}]
$\lambda$4363 emission line was detected at least at a 2$\sigma$ level,
allowing abundance determination by direct methods.
The oxygen abundance 12 + log O/H of the
SDSS galaxies lies in the range from $\sim$ 7.1 ($Z_\odot$/30) to
$\sim$ 8.5 (0.7 $Z_\odot$). The SDSS sample is merged with a sample of 109
blue compact dwarf (BCD) galaxies with high quality spectra, which contains
extremely  low-metallicity objects.  We use the merged sample to study the
abundance patterns of low-metallicity emission-line galaxies.
We find that extremely metal-poor galaxies (12 + log O/H $<$ 7.6, i.e. $Z$ $<$
$Z_\odot$/12)  are rare in the SDSS sample. The $\alpha$ element-to-oxygen
abundance ratios do not show
any significant trends with oxygen abundance, in agreement with previous
studies, except for a slight increase of Ne/O with increasing metallicity,
which we interpret as due to a moderate depletion of O onto grains in the most
metal-rich galaxies. The Fe/O abundance ratio is smaller than the solar value,
by up to 1 dex at the high metallicity end. We also find that Fe/O increases
with decreasing H$\beta$ equivalent width EW(H$\beta$). We interpret this as
a sign of strong depletion onto dust grains, and gradual destruction of those
grains on a time scale of a few Myr.
All the galaxies are found to have log  N/O $>$  --1.6, implying that they
have a different nature than the subsample of
high-redshift damped Ly$\alpha$ systems with
log N/O of $\sim$ --2.3 and that their ages are 
%probably 
larger than 100 -- 300 Myr. We confirm the apparent increase in N/O with 
decreasing
EW(H$\beta$), already shown in previous studies, and explain it as the
signature of gradual nitrogen ejection by massive stars from the most recent
starburst.
}
\titlerunning{The chemical composition of metal-poor emission line galaxies
in the DR3 of the SDSS}
\maketitle

%\markboth{Y.I.Izotov et al.}{Abundance patterns in the emission-line
%galaxies from the SDSS Early Data Release}

\section{Introduction}

Knowledge of the detailed chemical composition of galaxies is fundamental
to our understanding of stellar nucleosynthesis and evolution of galaxies.
Emission-line galaxies provide an easy way to determine the abundances of such
elements as He, N, O, Ne, S, Cl, Ar and Fe, out to redshifts of about 0.4,
from an
analysis of the radiation from their H {\sc ii} regions in the visible domain.
These determinations
are considered more reliable if the electron temperatures can be measured
directly, using the [O {\sc iii}] $\lambda$4363/[O {\sc iii}] $\lambda$5007
line ratio. Then, the ionic abundances can be derived directly from the
strengths of the emission lines. The
assumptions to derive elemental abundances are reduced to a minimum.
The weak [O {\sc iii}] $\lambda$4363 can only be measured in metal-poor
emission-line galaxies,  where the H {\sc ii} regions suffer little cooling
and are at high enough temperature to produce significant emission of
this emission line.
%[O {\sc iii}] $\lambda$4363.

Metal-poor galaxies are the least chemically evolved galaxies and thus provide
the simplest test beds for theories of chemical evolution of galaxies. They
also possibly constitute a local counterpart to primeval high-redshift
galaxies.
In this spirit, they have been the subject of many studies in the recent past
\citep[e.g.,][and references therein]{IT04b}. They have also been used for the
quest of the abundance of primordial helium \citep[e.g.][and references
therein]{IT04a}. Those studies were
based on limited samples of very different coverages and selection criteria.

The Sloan Digital Sky Survey \citep[SDSS,][]{Y00} offers a gigantic
data base of galaxies with well-defined selection criteria and observed in a
homogeneous way. In addition, the spectral resolution is much better than that
of most previous data bases on emission-line galaxies. From this data base,
it is possible to extract a
%complete
sample of emission-line galaxies
%according to
with well defined criteria.
%This will allow one to address the
%questions above on a sound statistical basis. This will also open the way for
%studies which, so far, have suffered strong limitation due to the paucity of
%data, such as the effect of environment on the evolution of galaxies, or
%variation of metallicities with redshift.

Two papers have already used the SDSS to find objects with detected
[O {\sc iii}] $\lambda$4363. The first one is by \citet{I04},
and was based on the Early Data Release. It studied the abundance patterns of
metal-poor emission-line galaxies, with emphasis on the N/O ratio. The second
one is by \citet{K04b}, who produced a catalogue of emission-line galaxies
with low oxygen abundances
from the Data Release 1. In both cases, the abundances were computed using
the formulae given by \citet{ITL94}.

Here, we use the Data Release 3 (DR3).
%, increasing by a factor of $\sim$ 5
%the number of galaxies examined by \citet{I04} and by a factor of $\sim$ 1.5
%the number of galaxies considered by \citet{K04b}. 
We adopted selection criteria different from those by
\citet{I04} and \citet{K04b} so that our sample is in general composed of 
objects with higher quality spectra and therefore more reliable abundances.
In the present paper we compute the
abundances with revised expressions, based on the most recent atomic
coefficients for the line emissivities and on appropriate series of
photoionization models for the ionization correction factors.
All the results are available in electronic tables. We believe that they
should provide a useful basis for many future works dealing with such objects.

The organization of the paper is as follows. The selection criteria used to
extract the objects and the resulting sample are described in section
\ref{sect2}.
The derivation of the elemental abundances is presented in
section \ref{sect3}.
The obtained abundance patterns are discussed
in section \ref{sect4}. The main conclusions of this study are presented in
section \ref{sect5}.

%%%%%%%%%%%%%%%%%%%%%%%%%%%%%%%%%%%%%%%%%%%%%%%%
%    Fig.1
%%%%%%%%%%%%%%%%%%%%%%%%%%%%%%%%%%%%%%%%%%%%%%%%
\begin{figure}[t]
\hspace*{-0.0cm}\psfig{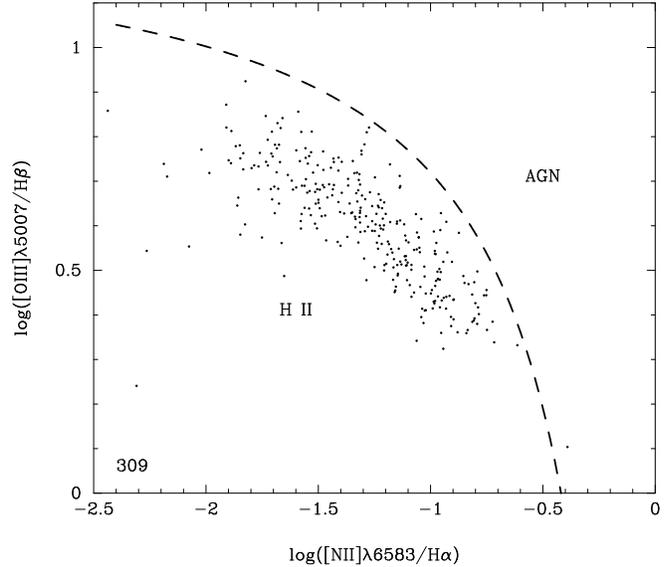}
\caption{Diagnostic diagram [O {\sc iii}] $\lambda$5007/H$\beta$ vs
[N {\sc ii}] $\lambda$6583/H$\alpha$ for our SDSS sample
(See Sect 2.2). The dashed line
separates galaxies with H {\sc ii} regions-like spectra from galaxies
containing an active nucleus (labeled ``AGN'') \citep{K03}. As in all
the figures with observational data, the number of
points is indicated in the left lower corner.}
\label{Fig1}
\end{figure}
%%%%%%%%%%%%%%%%%%%%%%%%%%%%%%%%%%%%%%%%%%%%%%%%%

%%%%%%%%%%%%%%%%%%%%%%%%%%%%%%%%%%%%%%%%%%%%%%%%
%    Fig.2
%%%%%%%%%%%%%%%%%%%%%%%%%%%%%%%%%%%%%%%%%%%%%%%%
\begin{figure}[t]
\hspace*{-0.0cm}\psfig{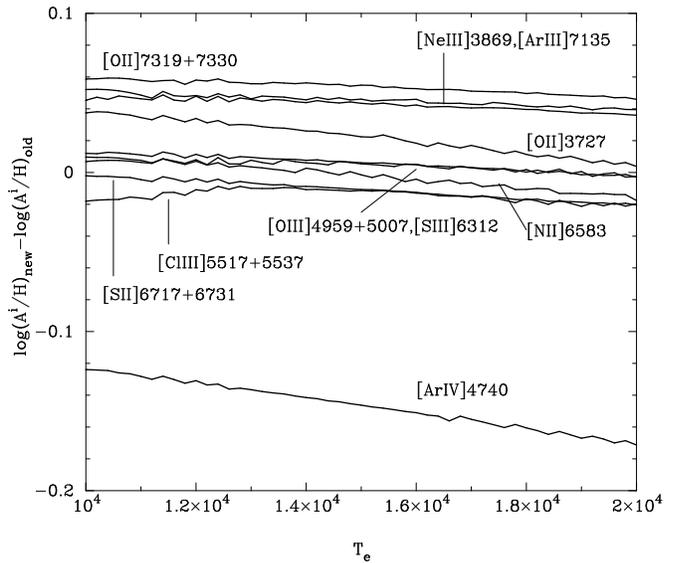}
\caption{Difference between ionic abundances calculated with the new
and old emissivities in the case of the low-density limit as a function
of the electron temperature. Curves are
labeled by the emission lines used for the abundance determination.}
\label{Fig2}
\end{figure}
%%%%%%%%%%%%%%%%%%%%%%%%%%%%%%%%%%%%%%%%%%%%%%%%%

%%%%%%%%%%%%%%%%%%%%%%%%%%%%%%%%%%%%%%%%%%%%%%%%
%    Fig.3
%%%%%%%%%%%%%%%%%%%%%%%%%%%%%%%%%%%%%%%%%%%%%%%%
\begin{figure*}[t]
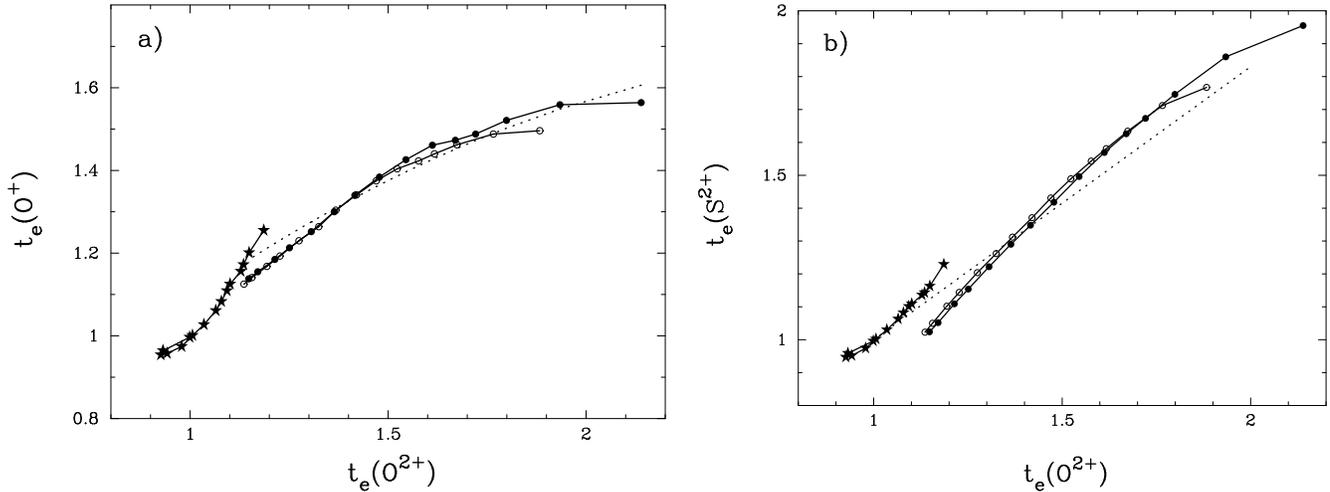

\hspace*{-0.0cm}\psfig{figure=3763f3a.ps,angle=270,height=6.5cm}%,clip=}
\hspace*{0.4cm}\psfig{figure=3763f3b.ps,angle=270,height=6.5cm,clip=}
\caption{Relations between the electron temperatures (a)
$t_{\rm e}$(O {\sc ii}) and $t_{\rm e}$(O {\sc iii})
($t_{\rm e}$ = 10$^{-4}$ $T_{\rm e}$) and (b)
$t_{\rm e}$(S {\sc iii}) and $t_{\rm e}$(O {\sc iii})
obtained from the sequences of photoionization models for H {\sc ii}
galaxies \citep{SI03} using stellar atmosphere models
from \citet{S02}. Filled circles are for the models
with $Z$ = 0.02 $Z_\odot$, open circles are for the models with
$Z$ = 0.05 $Z_\odot$, and stars are for the models
with $Z$ = 0.2 $Z_\odot$. The dotted lines are the approximation of the
models of \citet{S90} in (a) and the approximation of the
models of \citet{G92} in (b).}
\label{Fig3}
\end{figure*}
%%%%%%%%%%%%%%%%%%%%%%%%%%%%%%%%%%%%%%%%%%%%%%%%%

%%%%%%%%%%%%%%%%%%%%%%%%%%%%%%%%%%%%%%%%%%%%%%%%
%    Fig.4
%%%%%%%%%%%%%%%%%%%%%%%%%%%%%%%%%%%%%%%%%%%%%%%%
\begin{figure*}
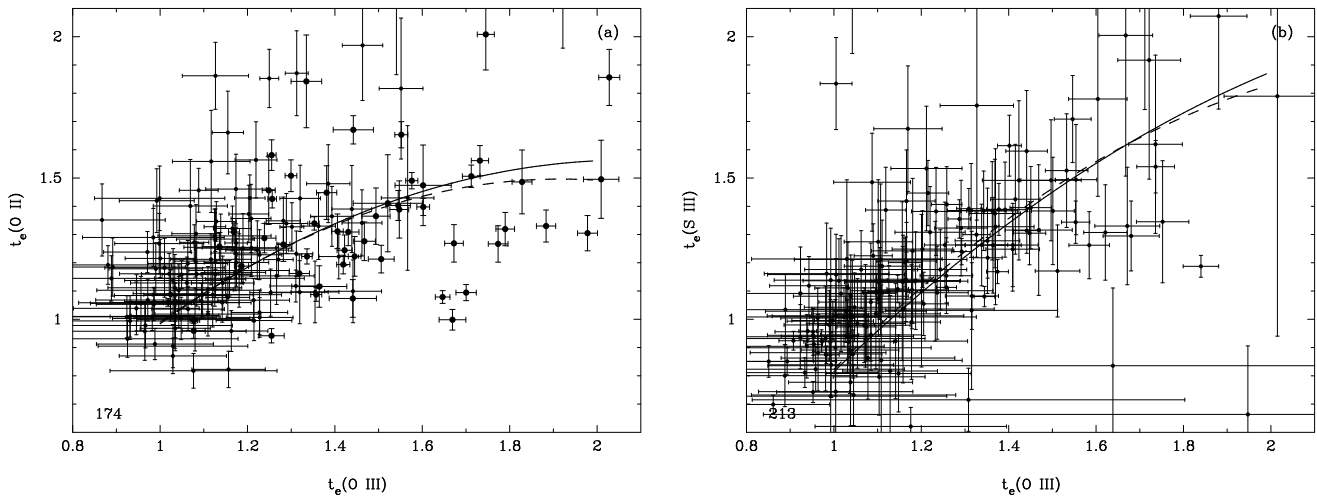

\hspace*{-0.0cm}\psfig{figure=3763f4a.ps,angle=270,height=6.5cm,clip=}
\hspace*{0.4cm}\psfig{figure=3763f4b.ps,angle=270,height=6.5cm,clip=}
\caption{(a) Relation between the electron temperatures
$t_{\rm e}$(O {\sc ii}) and $t_{\rm e}$(O {\sc iii}) for the SDSS sample
(small circles) and HeBCD galaxies (large circles).
(b) Relation between the electron temperatures
$t_{\rm e}$(S {\sc iii}) and $t_{\rm e}$(O {\sc iii}) for the SDSS
sample. The solid and dashed lines in (a) and (b) are
respectively the predictions for
H {\sc ii} region models (Eqs. \ref{toii} -- \ref{tsiii}) with low-
and intermediate metallicities.
}
\label{Fig4}
\end{figure*}
%%%%%%%%%%%%%%%%%%%%%%%%%%%%%%%%%%%%%%%%%%%%%%%%%

\section{The samples \label{sect2}}

\subsection{The SDSS DR3 data}

The SDSS DR3 \citep{A05} provides spectra in the wavelength range 
$\sim$ 3800 -- 9300\AA\ of $\sim$ 530\,000
galaxies, quasars and stars  selected over
4188 square degrees, and tables of measured parameters from these data.
We have extracted the flux-calibrated spectra of the SDSS DR3 galaxies
from the SDSS web page with the address http://www.sdss.org/dr3.

 From a visual examination of all the spectra and subsequent measurements
of emission line fluxes, we extracted $\sim$ 2700
spectra with an [O {\sc iii}] $\lambda$4363 emission line
detected at a level better than 1$\sigma$.
High-excitation Sy2 galaxies with strong [O {\sc iii}] $\lambda$4363
emission line can be recognized because of their strong
[O {\sc ii}]$\lambda$3727, He {\sc ii} $\lambda$4686,
[O {\sc iii}] $\lambda$5007 and [N {\sc ii}] $\lambda$6583 emission lines,
and have been removed from the sample.
The reduction of the spectra and the measurement of the line intensities is
done following the prescriptions of \citet{I04}. In particular, using the
IRAF/SPLOT routine, the line
intensities are fitted with a gaussian profile,
and the continuum is visually defined on
both sides of each line.

In the case of hydrogen lines, where the nebular emission line sits on
top of a much broader stellar absorption line, the level of the continuum is
set at the bottom of the absorption
feature. Given the spectral resolution of the SDSS, this absorption
feature is clearly seen whenever it is important. We estimate that, 
for the objects studied in this paper,  which have large equivalent 
widths of the
hydrogen emission lines, the
presence of this underlying absorption introduces only a small uncertainty
($<$ 10\%) on the fluxes of the strong hydrogen emission lines.
%The fraction of galaxies where the equivalent widths of
%hydrogen absorption lines are comparable to those of hydrogen emission lines
% does not exceed $\sim$ 10\%. The fraction of such H {\sc ii} regions in the
%restricted sample discussed later is smaller.
%*** Isuppressed this last sentence: it is not clear enough. besides 
%there are no more "restricted" samples, and samples A AND B ARE 
%DEFINED LATER"
%*** ARE YOU TALKING HERE OF OUR ENTIRE SDSS SAMPLE OR OF THE RESTRICTED
%SAMPLE? I SUPPOSE THAT FOR THE RESTRICTED SAMPLE, defined later, THIS
%FRACTION IS SMALLER. THIS SHOULD BE CLEARER IN THE TEXT***
%If the {\bf hydrogen} emission line is superposed on a
%{\bf stellar hydrogen} absorption
%line, the level of the continuum is set at the bottom of the absorption
%feature. {\bf The presence of absorption line introduces small uncertainty
%($\leq$ 10\%) on the fluxes of strong hydrogen emission line because
%the former line is much broader than the latter one. The fraction of the
%H {\sc ii} regions in our SDSS sample where the equivalent widths of
%hydrogen absorption lines are comparable to those of hydrogen emission lines
%is small and does not exceed $\sim$ 10\%.}

The ~3800 -- 9300\AA\ wavelength coverage of the SDSS spectra introduces
two sources of uncertainties in the emission line fluxes. First,
the emission line [O {\sc ii}] $\lambda$3727 in
the spectra of  galaxies with $z$ $\la$ 0.02 is not observed.
The fraction of such galaxies is $\sim$ 30\%.
In those cases the flux of the [O {\sc ii}] $\lambda$3727 emission
line is estimated from the fluxes of the
[O {\sc ii}] $\lambda$7320, 7330 emission lines taking into account the
empirical relation between the [O {\sc ii}] and [O {\sc iii}] temperatures
(see Eq. \ref{toii}).
\citet{K04b} have shown that the use of the [O {\sc ii}] $\lambda$7320, 7330 
emission lines instead of the [O {\sc ii}] $\lambda$3727 emission does not 
introduce biases in the derived oxygen abundances.
Therefore we use the former lines to derive the O$^+$ abundance in the
H {\sc ii} regions with the non-detected [O {\sc ii}] $\lambda$3727 emission
line. 
% in the spectra with non-observed
%[O {\sc ii}] $\lambda$3727 emission line.
Second, differential
refraction effects may be important. The majority of spectra in our
sample have been obtained at airmasses $\la$ 1.3. For an airmass of 1.3, the
offset of the H {\sc ii} region (or of a standard star) at wavelengths 4000\AA\
and 7500\AA\ relative to its nominal location inside the slit at 5000\AA\
are +0\farcs5 and $-$0\farcs5, respectively \citep{F82}. 
For comparison, the radius of the
round slit used in the spectroscopic observations is 1\farcs5. Therefore,
the observed fluxes of the blue lines [O {\sc ii}] $\lambda$3727 and
[Ne {\sc iii}] $\lambda$3868 and the red lines [Ar {\sc iii}] $\lambda$7135 and
[O {\sc ii}] $\lambda$7320, 7330 used for abundance determination could be
affected 
by differential refraction. From existing data we cannot estimate the
uncertainties introduced by this effect. Fortunately,
strong hydrogen emission lines in the wavelength range $\sim$ 3700 -- 6600\AA\
are present in the spectra of the majority of the selected galaxies. Correction
for extinction using the observed fluxes of these lines also corrects at 
the same time for the effects of differential refraction.

The precision of abundance determinations is critically dependent on
the uncertainties in the measurements of the [O {\sc iii}] $\lambda$4363
emission line fluxes. In the case of galaxies with weak
[O {\sc iii}] $\lambda$4363 emission, observational uncertainties \citep{I04}
may create spurious trends
in the abundance patterns. Therefore, we restricted our initial 
sample of 2700 objects
to objects with the best derived abundances.
  Empirically, we find that we can significantly reduce
unphysical trends when restricting the sample to
objects with an observed flux in the H$\beta$ emission line
larger than 10$^{-14}$ erg s$^{-1}$cm$^{-2}$.
Additionally, we have
excluded  all galaxies with both
[O {\sc iii}]$\lambda$4959/H$\beta$
$<$ 0.7 and [O {\sc ii}]$\lambda$3727/H$\beta$ $>$ 1.0. A weak
[O {\sc iii}] $\lambda$4363 line in these galaxies may be subject to
significant enhancement due to a mechanism different from  photoionization
heating in H {\sc ii} regions
\citep{I04} \footnote{The second condition
[O {\sc ii}]$\lambda$3727/H$\beta$ $>$ 1.0 ensures that,  by removing 
objects with weak [O {\sc iii}]$\lambda$4959 we do not remove at the 
same time  the
most metal-deficient high-excitation H {\sc ii} regions. For example, 
both the SE and NW components of I Zw 18
have [O {\sc iii}]$\lambda$4959/H$\beta$ $<$ 0.7, but also weak
[O {\sc ii}]$\lambda$3727/H$\beta$. Therefore, they are not removed 
from the sample.}.
Applying all these selection criteria gives us a sample of $\sim$
310 SDSS objects, consisting mainly of high-excitation H {\sc ii} regions.
%:  $\sim$ 1.5 times the number of galaxies in the catalog by
%\citet{K04b} and about 5 times as many as in the \citet{I04} sample.
The [O {\sc iii}] $\lambda$4363 emission line intensities in these 
galaxies is detected at the $\ga$ 2$\sigma$ level for all of them 
and at the $\ga$ 5$\sigma$ for half of them.

%Our sample is presented in Table \ref{Tab1}, which lists the plate names,
%the equatorial coordinates for the epoch J2000.0, the redshift $z$ and the
%SDSS $r$ apparent magnitude $m_r$. For galaxies which already have other
%names, those are given in the last column of Table \ref{Tab1}.
    The observed line fluxes $F$($\lambda$) are measured
using the IRAF/SPLOT routine.
The errors in the line fluxes are calculated from the photon statistics.
For this we use files attached to each spectrum which are  
generated by the pipeline data reduction and which contain 
the error in each pixel. Additionally, we adopt an error of 2\% in the
emission line fluxes to account for the uncertainties of standard 
star absolute
fluxes used for flux calibration \citep[e.g., ][]{O90}.
These errors will be later propagated into the calculation of abundance errors.
The line fluxes were corrected for both reddening \citep{W58}
and underlying hydrogen stellar absorption derived simultaneously by an
iterative procedure as described in \citet{ITL94}.
This procedure automatically corrects emission line fluxes for 
differential refraction effects.
The spectrum plate number in the SDSS database, the extinction-corrected emission line fluxes 
$I$($\lambda$) normalized to $I$(H$\beta$), together with their uncertainties,
the equivalent width EW, the observed flux $F$ of the H$\beta$ emission line
and the extinction coefficient $C$(H$\beta$) are shown in Table \ref{tab1}.

%\begin{table*}
%\caption{General characteristics of the SDSS galaxies$^1$ \label{Tab1}}
%\begin{tabular}{lrrrccl} \hline
%\#&Plate name    & R.A.(J2000)&Dec.(J2000) & $z$  & $m_r$ & Other names \\ \hline
% 1&0266-100& 09:44:01.86&   --00:38:32.18&0.0048&17.49&CGCG 007--025 \\
% 2&0266-155& 09:45:17.57&   --00:01:47.96&0.0217&19.16&SDSS J094517.56$-$000148.0 \\
% 3&0267-384& 09:49:54.16&     00:36:58.65&0.0063&17.68&Mrk 1236 \\
% 4&0267-421& 09:50:23.32&     00:42:29.28&0.0976&18.90&SDSS J095023.32$+$004229.2 \\
% 5&0268-200& 09:58:30.25&     00:02:42.98&0.0065&19.90&SDSS J095830.26$+$000243.0 \\ \hline
%0269-356& 10:02:23.39&     00:15:15.53&0.0437&18.89&SDSS J100223.39$+$001515.5 \\
%0270-306& 10:08:35.95&   --00:40:36.65&0.0213&19.03&SDSS J100835.95$-$004036.5 \\ \hline
%\end{tabular}

%$^1$Only first five galaxies are shown. The total list of galaxies
%is available in the electronic form at the Strasbourg database service.
%\end{table*}

\begin{table*}
\caption{Emission line fluxes in the SDSS galaxies$^1$ \label{tab1}}
\begin{tabular}{lccccccccccrrc} \hline
Spectrum    &[Ne{\sc iii}]&[O{\sc iii}]&[Fe{\sc iii}]
            &[O{\sc iii}]&He {\sc i}&[O {\sc i}]&[S{\sc iii}]&[N{\sc 
ii}]&[S{\sc ii}]
         &[Ar{\sc iii}]&EW&$F^2$&$C$(H$\beta$) \\
number      & 3869  & 4363  & 4658  & 4959  & 5876  & 6300  & 6312  & 6584  & 6725  & 7135  &\multicolumn{1}{c}{\AA}&& \\ \hline
0266-51630-100$^3$& 0.370 & 0.116 & 0.005 & 1.911 & 0.117 & 0.024 & 0.017 & 0.036 & 0.142 & 0.055 & 262&  502&  0.00  \\
                  & 0.012 & 0.005 & 0.002 & 0.057 & 0.004 & 0.002 & 0.001 & 0.002 & 0.004 & 0.003 &    &     &        \\
0267-51608-384$^3$& 0.444 & 0.042 & 0.010 & 1.263 & 0.088 & 0.020 & 0.010 & 0.096 & 0.219 & 0.053 &  57&  314&  0.19 \\
                  & 0.018 & 0.006 & 0.004 & 0.042 & 0.005 & 0.003 & 0.003 & 0.005 & 0.008 & 0.003 &    &     &        \\
0270-51909-306$^3$& 0.175 & 0.015 &  ...  & 0.967 & 0.130 & 0.026 & 0.011 & 0.294 & 0.518 &  ...  &  55&  104&  0.10 \\
                  & 0.014 & 0.009 &  ...  & 0.037 & 0.010 & 0.006 & 0.006 & 0.015 & 0.019 &  ...  &    &     &        \\
0272-51941-365$^3$& 0.297 & 0.023 & 0.006 & 1.131 & 0.112 & 0.058 & 0.023 & 0.224 & 0.502 & 0.076 &  31&  110&  0.13  \\
                  & 0.019 & 0.012 & 0.011 & 0.042 & 0.010 & 0.009 & 0.008 & 0.013 & 0.020 & 0.008 &    &     &        \\
0275-51910-429$^3$& 0.430 & 0.094 &  ...  & 1.587 & 0.112 & 0.029 & 0.012 & 0.059 & 0.237 & 0.056 & 107&  197&  0.03  \\
                  & 0.017 & 0.007 &  ...  & 0.051 & 0.006 & 0.004 & 0.003 & 0.004 & 0.009 & 0.004 &    &     &        \\
%0269-356$^3$& 2.14   & 0.157 & 0.023 & 0.019 & 0.735 & 0.018 & 0.016 & 0.308 & 0.688 & 0.087 & 0.034 &  46&   74&  0.085 \\
%        & 0.05   & 0.007 & 0.007 & 0.005 & 0.018 & 0.005 & 0.007 & 0.011 & 0.017 & 0.008 & 0.011 &    &     &        \\
%0270-306$^3$&  ...   & 0.175 & 0.015 &  ...  & 0.967 & 0.013 & 0.011 & 0.294 & 0.518 &  ...  & 0.030 &  55&  104&  0.100 \\
%        &  ...   & 0.006 & 0.003 &  ...  & 0.019 & 0.004 & 0.007 & 0.009 & 0.011 &  ...  & 0.006 &    &     &        \\
\hline
\end{tabular}

$^1$Only the first five galaxies are shown here. The total list of galaxies
is available in electronic form.% at the Strasbourg database service.
The hydrogen, [O {\sc ii}] $\lambda$3727, He {\sc ii} $\lambda$4686, [Ar {\sc iv}] $\lambda$4740, [Fe {\sc iii}] $\lambda$4988,
[Cl {\sc iii}] $\lambda$5517,5531, [O {\sc ii}] $\lambda$7320,7330
and [S {\sc iii}] $\lambda$9069
emission lines are not shown here for lack of space,
but they are present in the electronic table.

%$^2$in \AA.

$^2$in units 10$^{-16}$ erg s$^{-1}$ cm$^{-2}$.

$^3$The first row are values, the second row are errors.

\end{table*}

Figure \ref{Fig1} shows the location of the galaxies of the SDSS sample in the 
%classical
[O {\sc
iii}] $\lambda$5007/H$\beta$ vs
[N {\sc ii}] $\lambda$6583/H$\alpha$ diagram 
which is generally used to distinguish H {\sc ii}
galaxies ionized only by massive stars from galaxies containing an active
nucleus (AGN). The dashed line separates the two classes of 
galaxies \citep{K03}. As in \citet{I04}, all but one of
our selected objects  fall in the ``H II'' region of the diagram. 
%This comes from the fact that we impose the detection of [O {\sc iii}]
%$\lambda$4363 emission line, which implies that overall the galaxies are
%metal-poor while galaxies containing an active nucleus are generally
%metal-rich.

%%%%%%%%%%%%%%%%%%%%%%%%%%%%%%%%%%%%%%%%%%%%%%%%
%    Fig.5
%%%%%%%%%%%%%%%%%%%%%%%%%%%%%%%%%%%%%%%%%%%%%%%%
\begin{figure}
\hspace*{-0.0cm}\psfig{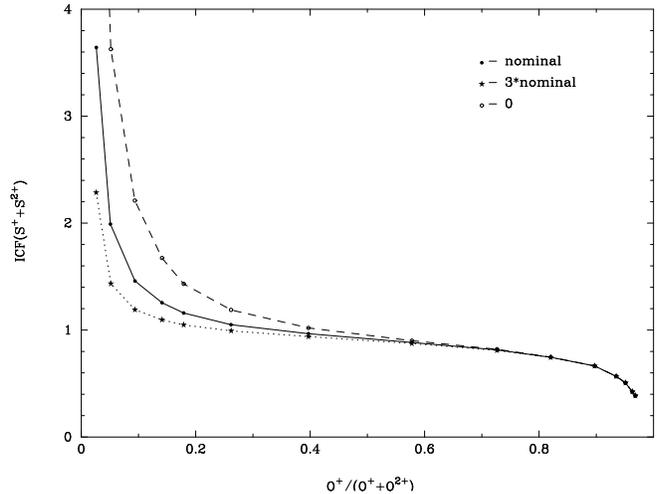}
\caption{Ionization correction factor $ICF$(S$^+$+S$^{2+}$) vs the fraction
O$^+$/(O$^+$+O$^{2+}$) for the ``low $Z$'' models with different rates of
dielectronic recombination for sulfur ions. Nominal dielectronic recombination
for sulfur ions is the same as that for oxygen ions.}
\label{Fig5}
\end{figure}
%%%%%%%%%%%%%%%%%%%%%%%%%%%%%%%%%%%%%%%%%%%%%%%%%

\subsection{The HeBCD sample}

In addition to the SDSS sample, we also use a sample of galaxies
which has been collected primarily to study the helium abundances in
low-metallicity blue compact dwarf galaxies (the HeBCD sample).
This sample is the same as in \citet{I04},
with the addition of galaxies from \citet{IT04a}.

%%%%%%%%%%%%%%%%%%%%%%%%%%%%%%%%%%%%%%%%%%%%%%%%
%    Fig.6
%%%%%%%%%%%%%%%%%%%%%%%%%%%%%%%%%%%%%%%%%%%%%%%%
\begin{figure*}
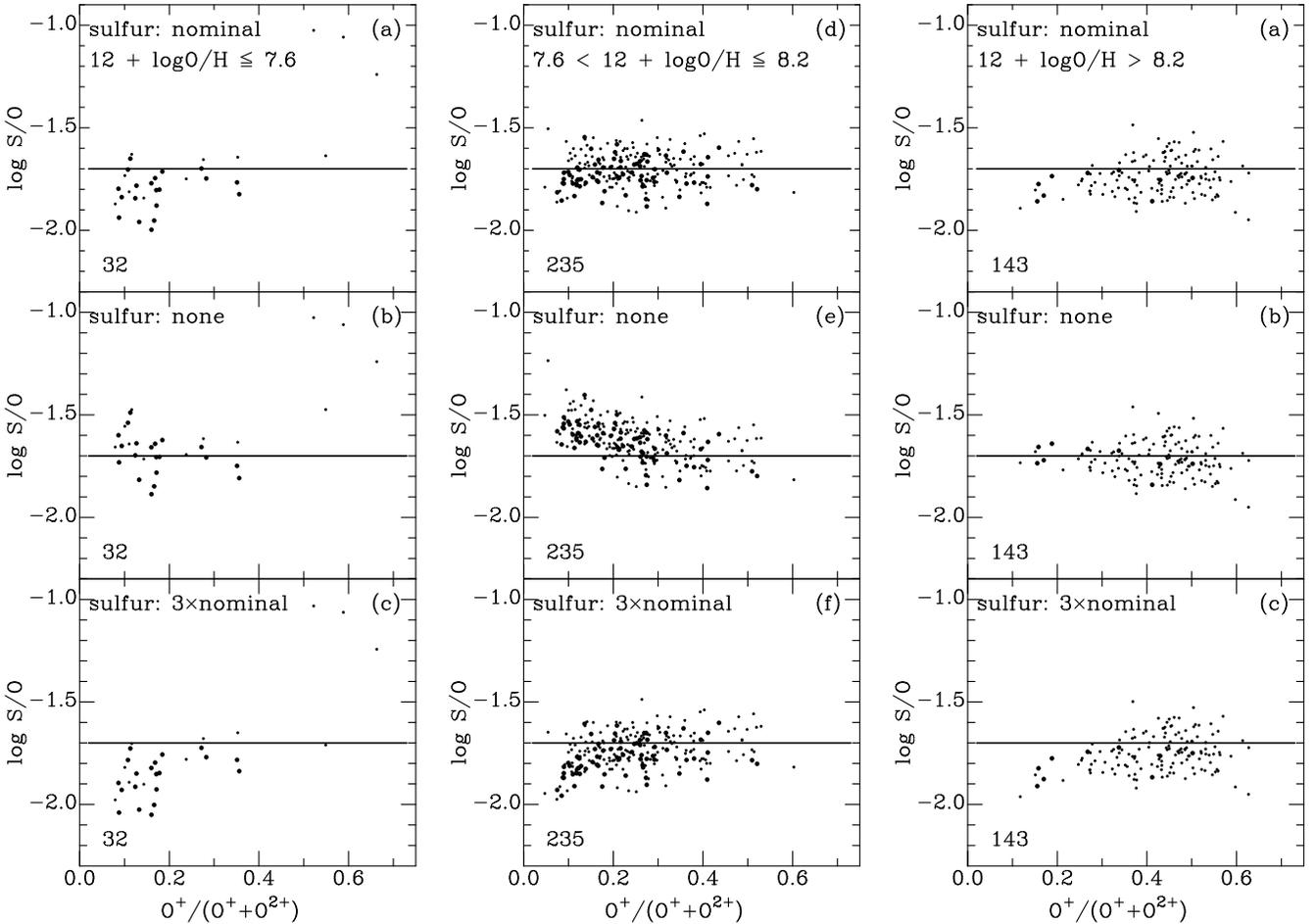

\hspace*{-0.0cm}\psfig{figure=3763f6a.ps,angle=0,height=12.5cm}%,clip=}
\hspace*{0.4cm}\psfig{figure=3763f6b.ps,angle=0,height=12.5cm}%,clip=}
\hspace*{0.4cm}\psfig{figure=3763f6c.ps,angle=0,height=12.5cm,clip=}
\caption{{\bf left} Sulfur-to-oxygen abundance ratio vs the fraction
O$^+$/(O$^+$+O$^{2+}$) in the H {\sc ii} regions with 12 + log O/H $\leq$ 7.6
for different rates of dielectronic recombination for sulfur.
The nominal dielectronic recombination for S ions is the same as that 
for O ions.
Large filled circles are data for the HeBCD sample, dots are for
the SDSS sample.
Solid horizontal line shows the mean S/O abundance ratio from
the data with the nominal rate of dielectronic recombination for sulfur ions.
{\bf middle} The same as in the left panel but for H {\sc ii} regions with
7.6 $<$ 12 + log O/H $\leq$ 8.2. {\bf right} The same as in the left panel but
for H {\sc ii} regions with 12 + log O/H $>$ 8.2. The total number of 
galaxies in each panel is swown in the lower left corner.}
\label{Fig6}
\end{figure*}
%%%%%%%%%%%%%%%%%%%%%%%%%%%%%%%%%%%%%%%%%%%%%%%%%

%%%%%%%%%%%%%%%%%%%%%%%%%%%%%%%%%%%%%%%%%%%%%%%%
%    Fig.7
%%%%%%%%%%%%%%%%%%%%%%%%%%%%%%%%%%%%%%%%%%%%%%%%
\begin{figure*}
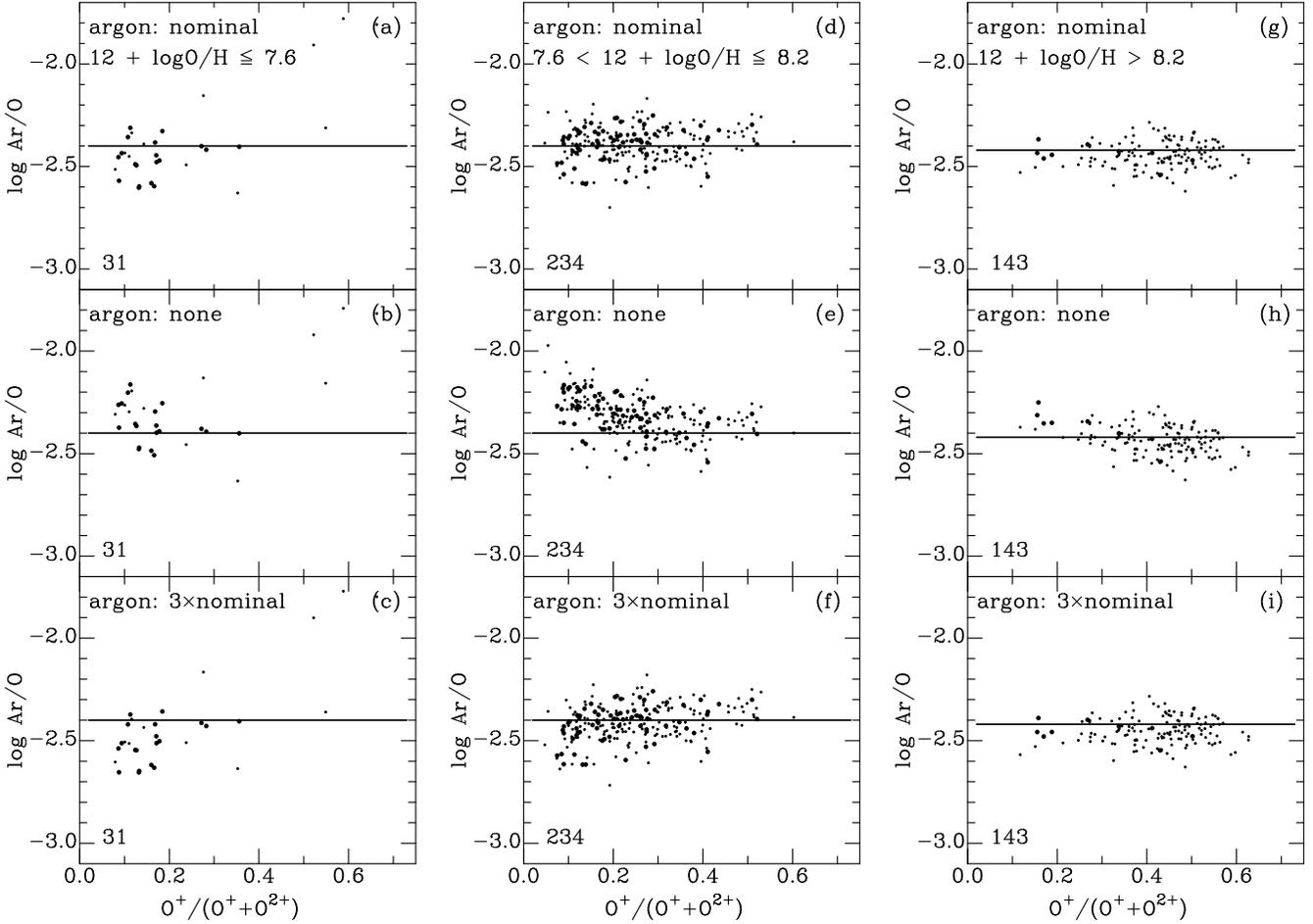

\hspace*{-0.0cm}\psfig{figure=3763f7a.ps,angle=0,height=12.5cm}%,clip=}
\hspace*{0.4cm}\psfig{figure=3763f7b.ps,angle=0,height=12.5cm}%,clip=}
\hspace*{0.4cm}\psfig{figure=3763f7c.ps,angle=0,height=12.5cm,clip=}
\caption{The same as in Fig. \ref{Fig6} but for argon.
%{\bf left} Argon-to-oxygen abundance ratio vs the fraction
%O$^+$/(O$^+$+O$^{2+}$) in the H {\sc ii} regions with 12 + log O/H $\leq$ 7.6
%for different rates of dielectronic recombination for argon.
Nominal dielectronic recombination for Ar ions is the same as that for Ne ions.
%Large filled circles are data for the HeBCD sample, dots are for the
%SDSS data.
%{\bf middle} The same as in the left panel but for H {\sc ii} regions with
%7.6 $<$ 12 + log O/H $\leq$ 8.2. {\bf right} The same as in the left panel but
%for H {\sc ii} regions with 12 + log O/H $>$ 8.2.
}
\label{Fig7}
\end{figure*}
%%%%%%%%%%%%%%%%%%%%%%%%%%%%%%%%%%%%%%%%%%%%%%%%%

%%%%%%%%%%%%%%%%%%%%%%%%%%%%%%%%%%%%%%%%%%%%%%%%
%    Fig.8
%%%%%%%%%%%%%%%%%%%%%%%%%%%%%%%%%%%%%%%%%%%%%%%%
\begin{figure*}
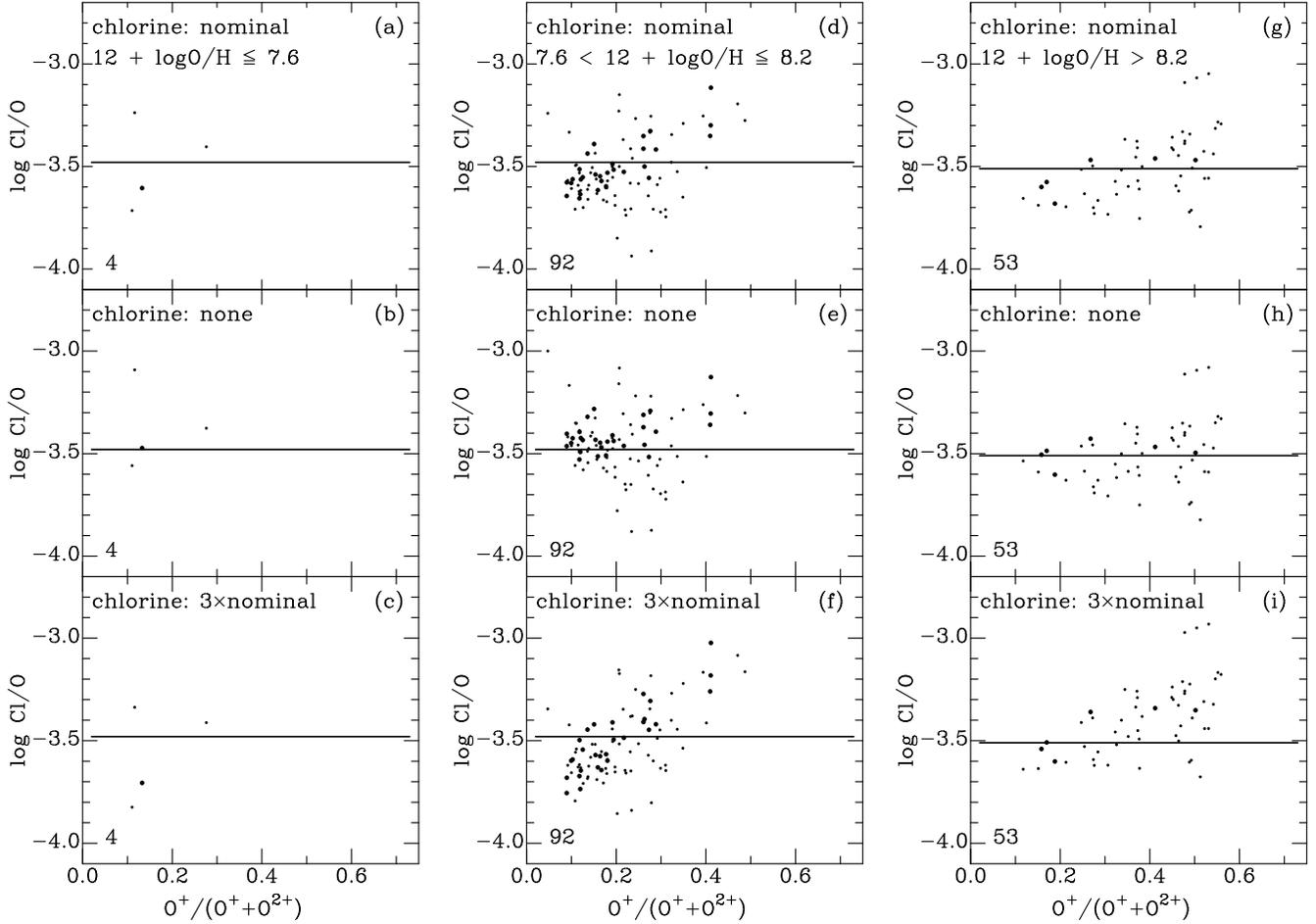

\hspace*{-0.0cm}\psfig{figure=3763f8a.ps,angle=0,height=12.5cm}%,clip=}
\hspace*{0.4cm}\psfig{figure=3763f8b.ps,angle=0,height=12.5cm}%,clip=}
\hspace*{0.4cm}\psfig{figure=3763f8c.ps,angle=0,height=12.5cm,clip=}
\caption{The same as in Fig. \ref{Fig6} but for chlorine.
%{\bf left} Chlorine-to-oxygen abundance ratio vs the fraction
%O$^+$/(O$^+$+O$^{2+}$) in the H {\sc ii} regions with 12 + log O/H $\leq$ 7.6
%for different rates of dielectronic recombination for chlorine.
Nominal dielectronic recombination for Cl ions is the same as that for O ions.
%Large filled circles are data for the HeBCD sample, dots are for the
%SDSS data.
%{\bf middle} The same as in the left panel but for H {\sc ii} regions with
%7.6 $<$ 12 + log O/H $\leq$ 8.2. {\bf right} The same as in the left panel but
%for H {\sc ii} regions with 12 + log O/H $>$ 8.2.
}
\label{Fig8}
\end{figure*}
%%%%%%%%%%%%%%%%%%%%%%%%%%%%%%%%%%%%%%%%%%%%%%%%%

%%%%%%%%%%%%%%%%%%%%%%%%%%%%%%%%%%%%%%%%%%%%%%%%
%    Fig.9
%%%%%%%%%%%%%%%%%%%%%%%%%%%%%%%%%%%%%%%%%%%%%%%%
\begin{figure*}
\hspace*{-0.0cm}\psfig{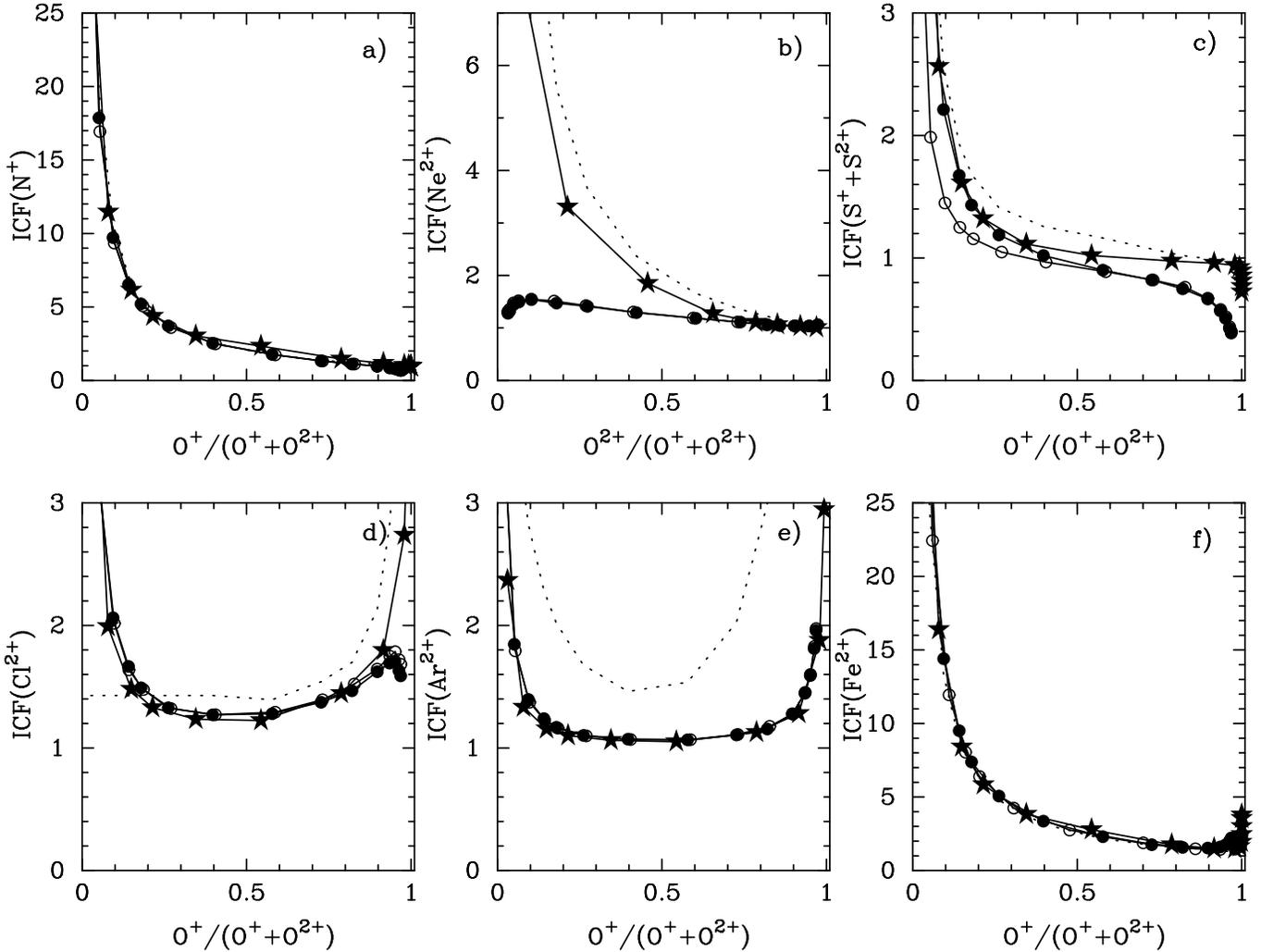}
\caption{Ionization correction factors for heavy elements.
Filled circles are for the models
with $Z$ = 0.02 $Z_\odot$, open circles are for the models with
$Z$ = 0.05 $Z_\odot$, and stars are for the models
with $Z$ = 0.2 $Z_\odot$. The dotted line is the approximation of the
models of \citet{S90}. }
\label{Fig9}
\end{figure*}
%%%%%%%%%%%%%%%%%%%%%%%%%%%%%%%%%%%%%%%%%%%%%%%%%

%%%%%%%%%%%%%%%%%%%%%%%%%%%%%%%%%%%%%%%%%%%%%%%%%

\section{Element abundance determination \label{sect3}}

The physical conditions and element abundances are derived from the
reddening-corrected emission line fluxes using the electron temperature
derived from the  [O {\sc iii}] $\lambda$4363/($\lambda$4959+$\lambda$5007)
ratio. Temperature corrections to derive ionic abundances and ionization
corrections to derive elemental abundances are based on sequences of
photoionization models that best reproduce the observed emission line trends
in H {\sc ii} galaxies \citep{SI03}.

\subsection{Atomic data and ionic abundances}

In previous studies, we have used emissivities for emission lines provided by
\citet{A84}. Since that time the collisional strengths for
many transitions have been re-evaluated. Here, we use the 
atomic data from the references  listed in \citet{S05}.

The electron temperatures derived from the
[O {\sc iii}]$\lambda$4363/$\lambda$5007 line ratio using old and new
emissivities differ by less than
1\% in the temperature range 5000--20000K. Therefore, we derive
the electron temperature from an iterative procedure, using the
equation from \citet{A84}:
\begin{equation}
t=\frac{1.432}{\log [(\lambda 4959+\lambda 5007)/\lambda 4363]-\log C_T},
\label{toiii}
\end{equation}
where $t=10^{-4}T_e$(O {\sc iii}),
\begin{equation}
C_T=(8.44-1.09t+0.5t^2-0.08t^3)\frac{1+0.0004x}{1+0.044x},\nonumber
\end{equation}
and $x$=10$^{-4}$$N_e$$t^{-0.5}$.
The density  $N_e$(S {\sc ii}) is derived
by interpolation of the
[S {\sc ii}]$\lambda$6717/$\lambda$6731 line ratio vs $N_e$(S {\sc ii}) 
relation)
with the use of the collisional strengths
from \citet{R96}.
%$N_e$(S {\sc ii}) is always smaller
%than 10$^3$ cm$^{-3}$, so that the term containing $x$ is never important.
The derived $N_e$ is always smaller
than 10$^3$ cm$^{-3}$, so that the term containing $x$ is never important.

In Fig. \ref{Fig2} we show the differences between the 
ionic abundances calculated
with new and old atomic data as a function of the
electron temperature. Each curve is labeled with the emission line used for
the ionic abundance determination.
We use [N {\sc ii}] $\lambda$6548, 6584 for the determination of the nitrogen
abundance, [O {\sc ii}]
$\lambda$3727 (or [O {\sc ii}] $\lambda$7320,7331 when
%the latter
[O {\sc ii}] $\lambda$3727 is not available) and
[O {\sc iii}] $\lambda$4959, 5007 for the oxygen abundance, [Ne {\sc iii}]
$\lambda$3868 for the neon abundance, [S {\sc iii}] $\lambda$6312 and
[S {\sc ii}] $\lambda$6717, 6731 for the
sulfur abundance when detected,
[Cl {\sc iii}] $\lambda$5517, 5531 for the chlorine abundance, [Ar {\sc iii}]
$\lambda$7135
(and [Ar {\sc iv}] $\lambda$4740 if seen)
for the argon
abundance, and [Fe {\sc iii}] $\lambda$4658 for the
iron abundance. In general, the new atomic data change the ionic abundances by
not more than 0.06 dex. The only exception is Ar$^{3+}$ where the new
determinations are more than 0.1 dex lower than the old ones.
Following \citet{S05} we do not attempt to analyze the collisional strengths
for Fe$^{2+}$ and use the same ones as in \citet{TIL95}.
We obtain linear fits for the dependences shown in Fig. \ref{Fig2} and use
these fits in the abundance determinations. The ionic abundances
are then derived from the equations:
\begin{eqnarray}
12+\log {\rm O}^{+}/{\rm H}^+=&\log \frac{\lambda 3727}{{\rm 
H}\beta}+5.961+\frac{1.676}{t}
-0.40\log t \nonumber\\
                  &-0.034t+\log (1+1.35x),
\label{oii1}
\end{eqnarray}
\begin{eqnarray}
12+\log {\rm O}^{+}/{\rm H}^+=&\log \frac{\lambda 7320+\lambda 
7330}{{\rm H}\beta}+6.901 \nonumber\\
                               &+\frac{2.487}{t}-0.483\log t \nonumber\\
                  &-0.013t+\log (1-3.48x), %\nonumber\\
\label{oii2}
\end{eqnarray}
\begin{eqnarray}
12+\log {\rm O}^{2+}/{\rm H}^+=&\log \frac{\lambda 4959+\lambda 
5007}{{\rm H}\beta}+6.200+\frac{1.251}{t}
  \nonumber\\
                  &-0.55\log t-0.014t,
\label{oiii}
\end{eqnarray}
\begin{eqnarray}
12+\log {\rm N}^{+}/{\rm H}^+=&\log \frac{\lambda 6548+\lambda 
6584}{{\rm H}\beta}+6.234 \nonumber\\
                               &+\frac{0.950}{t}-0.42\log t \nonumber\\
                               &-0.027t+\log (1+0.116x),
\label{nii}
\end{eqnarray}
\begin{eqnarray}
12+\log {\rm Ne}^{2+}/{\rm H}^+=&\log \frac{\lambda 3869}{{\rm 
H}\beta}+6.444+\frac{1.606}{t}
  \nonumber\\
                  &-0.42\log t-0.009t,
\label{neiii}
\end{eqnarray}
\begin{eqnarray}
12+\log {\rm S}^{+}/{\rm H}^+=&\log \frac{\lambda 6717+\lambda 
6731}{{\rm H}\beta}+5.439 \nonumber\\
                               &+\frac{0.929}{t}-0.28\log t \nonumber\\
                               &-0.018t+\log (1+1.39x),
\label{sii1}
\end{eqnarray}
\begin{eqnarray}
12+\log {\rm S}^{2+}/{\rm H}^+=&\log \frac{\lambda 6312}{{\rm 
H}\beta}+6.690+\frac{1.678}{t}
  \nonumber\\
                  &-0.47\log t-0.010t,
\label{sii2}
\end{eqnarray}
\begin{eqnarray}
12+\log {\rm Cl}^{2+}/{\rm H}^+=&\log \frac{\lambda 5517+\lambda 
5537}{{\rm H}\beta}+5.700+\frac{1.088}{t}
  \nonumber\\
                  &-0.483\log t-0.002t,
\label{cliii}
\end{eqnarray}
\begin{eqnarray}
12+\log {\rm Ar}^{2+}/{\rm H}^+=&\log \frac{\lambda 7135}{{\rm H}\beta}+6.174
  \nonumber\\
                                 &+\frac{0.799}{t}-0.48\log t \nonumber\\
                                 &-0.013t+\log (1+0.22x), \nonumber\\
\label{ariii1}
\end{eqnarray}
\begin{eqnarray}
12+\log {\rm Ar}^{3+}/{\rm H}^+=&\log \frac{\lambda 4740}{{\rm 
H}\beta}+6.308+\frac{1.280}{t}
  \nonumber\\
                  &-0.48\log t-0.047t,
\label{ariii2}
\end{eqnarray}
\begin{eqnarray}
12+\log {\rm Fe}^{2+}/{\rm H}^+=&\log \frac{\lambda 4658}{{\rm 
H}\beta}+6.498+\frac{1.298}{t}
  \nonumber\\
                  &-0.48\log t.
\label{feiii}
\end{eqnarray}
%\begin{eqnarray}
%12+\log {\rm Fe}^{2+}/{\rm H}^+=&\log \frac{\lambda 4988}{{\rm H}\beta}+6.498+\frac{1.298}{t}
% \nonumber\\
%                 &-0.48\log t.
%\label{feiii1}
%\end{eqnarray}

Since different ions reside in different parts of the H {\sc ii} region,
the electron temperatures in Eq. \ref{oii1} - \ref{feiii} are not the
same for the various ions. Only
$T_{\rm e}$(O {\sc iii}) is derived directly from observations for all
galaxies.
Otherwise, we use the relations between
$T_{\rm e}$(O {\sc iii}) and the temperatures characteristic of other ions in
sequences of photoionization models that best fit the observed line emission
trends in H {\sc ii} galaxies. These sequences are defined as in
\citet{SI03}, but have been recomputed with the new atomic data, and with an
input radiation field computed with Starburst 99 \citep{L99} using
the stellar model atmospheres described in \citet{S02}.
We thus obtain
the relations between the electron temperatures $T_e$(O {\sc iii}) and
$T_e$(O {\sc ii})
which are shown by solid lines in Fig. \ref{Fig3}a.
Adopting the nomenclature
of \citet{SI03}, ``low $Z$'' models are represented by filled circles,
``intermediate $Z$'' models by open circles and
``high $Z$'' models by stars. These $Z$ correspond
to oxygen abundances 12 + log O/H = 7.2,
7.6 and 8.2, respectively. For comparison, the dotted line is the
fit of \citet{ITL94,ITL97} to the earlier models of \citet{S90}.
The latter approximation, used in all our
previous studies, is not very different from the new relations shown
by solid lines in Fig. \ref{Fig3}.
We approximate the relations from the new models  by the expressions
\begin{eqnarray}
%t({\rm O\ II})=&-0.647+t\times(2.118-0.490t),& {\rm low\ }Z, \nonumber\\
%               &-0.862+t\times(2.453-0.621t),& {\rm intermed.\ }Z, \nonumber\\
%               &-0.253+t\times(1.432-0.182t),& {\rm high\ }Z, \label{toii}
%t({\rm O\ II})=&-0.577+2.065t-0.498t^2,& {\rm log O/H}=-4.8, \nonumber\\
%               &-0.744+2.338t-0.610t^2,& {\rm log O/H}=-4.4, \nonumber\\
%               &2.967-4.797t+2.827t^2,& {\rm log O/H}=-3.8, \nonumber\\
%\label{toii}
t({\rm O\ II})&=-0.577+t\times(2.065-0.498t),& {\rm low\ }Z, \nonumber\\
               &=-0.744+t\times(2.338-0.610t),& {\rm intermed.\ }Z, \nonumber\\
               &=2.967+t\times(-4.797+2.827t),& {\rm high\ }Z, \label{toii}
\end{eqnarray}
where $t$(O {\sc ii}) = 10$^{-4}$$T$(O {\sc ii}) and
$t$ = 10$^{-4}$$T$(O {\sc iii}).

In the H {\sc ii} regions where both [O {\sc ii}] $\lambda$3727 and
[O {\sc ii}] $\lambda$7320, 7331 are detected, the electron temperature
$T_{\rm e}$(O {\sc ii}) can be derived directly from the
$\lambda$3727/($\lambda$7320+$\lambda$7331) ratio. In Fig. \ref{Fig4}a we
compare the relations between $T_{\rm e}$(O {\sc ii})
and $T_{\rm e}$(O {\sc iii}) obtained from the observations
with those from
model predictions. Large filled circles correspond to galaxies from the
HeBCD sample, while dots correspond to SDSS galaxies. Despite a large
scatter of the data points caused mainly by large flux errors of the weak
[O {\sc ii}] $\lambda$7320, 7331 emission
lines, the relation between $T_{\rm e}$(O {\sc ii}) and
$T_{\rm e}$(O {\sc iii}) derived from the observations follows generally the
one obtained from the models. Therefore, in the following we adopt the
model relation to derive $T_{\rm e}$(O {\sc ii}). Note
that the scatter in Fig. \ref{Fig4} implies that, for the low redshift
objects of the SDSS sample  for which  [O {\sc ii}] $\lambda$3727 is outside
the observed wavelength range, the O$^+$ abundance is not very accurate
since it is obtained from  [O {\sc ii}] $\lambda$7320,7331.
This concerns about 30\% of the sample.

Similar relations between $T_{\rm e}$(S {\sc iii})  and
$T_{\rm e}$(O {\sc iii}) can be obtained from the same photoionization models:
\begin{eqnarray}
%t({\rm S\ III})=&-1.085+2.320t-0.420t^2,& {\rm log O/H}=-4.8, \nonumber\\
%               &-1.276+2.645t-0.546t^2,&  {\rm log O/H}=-4.4, \nonumber\\
%               &1.653-2.261t+1.605t^2,& {\rm log O/H}=-3.8, \nonumber\\
%\label{tsiii}
t({\rm S\ III})&=-1.085+t\times(2.320-0.420t),& {\rm low\ }Z, \nonumber\\
                &=-1.276+t\times(2.645-0.546t),& {\rm intermed.\ }Z, \nonumber\\
                &=1.653+t\times(-2.261+1.605t),& {\rm high\ }Z, \label{tsiii}
\end{eqnarray}
where $t$(S {\sc iii}) = 10$^{-4}$$T_{\rm e}$(S {\sc iii}) and
$t$ = 10$^{-4}$$T_{\rm e}$(O {\sc iii}).
They are shown in Fig. \ref{Fig3}b. The large wavelength coverage of the SDSS
spectra allows us to detect at the same time
the auroral [S {\sc iii}] $\lambda$6312 and nebular
[S {\sc iii}] $\lambda$9069 emission lines in low redshift galaxies  and
to derive the electron temperature $T_{\rm e}$(S {\sc iii}).
In Fig. \ref{Fig4}b we show the dependence of $T_{\rm e}$(S {\sc iii}) on
$T_{\rm e}$(O {\sc iii}) for those galaxies.
  Again, the theoretical relation is in satisfactory agreement with the
observations, if we take into account the large
scatter in the data, mostly due to measurement errors of the
weak [S {\sc iii}] $\lambda$6312 line.

We thus adopt  $T_e$(O {\sc ii}) from Eq. \ref{toii}
for the calculation of N$^+$, O$^+$, S$^+$ and Fe$^{2+}$ abundances,
$T_e$(S {\sc iii}) from Eq. \ref{tsiii} for the calculation
of S$^{2+}$, Cl$^{2+}$ and Ar$^{2+}$ abundances
and $T_e$(O {\sc iii}) from Eq. \ref{toiii} for the calculation of
O$^{2+}$ and Ne$^{2+}$
% and Ar$^{3+}$
abundances.
We adopt the ``low $Z$'' fits for H {\sc ii} regions
with 12 + log O/H $\leq$ 7.2, the ``high $Z$'' fits for the
H {\sc ii} regions with 12 + log O/H $\geq$ 8.2 (Eqs. \ref{toii} and
\ref{tsiii}), and linearly interpolate $T_e$(O {\sc ii}) and
$T_e$(S {\sc iii}) between ``low $Z$'' and ``intermediate $Z$'' cases
for H {\sc ii} regions with
12+log O/H between 7.2 and 7.6, and between ``intermediate $Z$'' and
``high $Z$'' cases for H {\sc ii} regions with 12+log O/H between 7.6 and 8.2.
Since  the population of the Ar$^{3+}$ ion
is generally smaller than that of the Ar$^{2+}$ ion and the
[Ar {\sc iv}] lines weaker than the [Ar {\sc iii}] lines, we decided to derive
the Ar abundance only from the abundance of the  Ar$^{2+}$ ion.

%The quality of the spectra is good enough to derive the
%N, O, Ne, S, Ar and Fe abundances.

\subsection{Ionization correction factors}

The total element abundance can be derived from the abundances of ions seen
in the optical spectra using ionization correction factors ($ICF$s). For this,
the ionization structure of the H {\sc ii} region has to be derived from
photoionization models. Fortunately, the most abundant ions of
oxygen, O$^+$ and O$^{2+}$, are seen in the optical spectra of the H {\sc ii}
regions. This fact allows one to immediately derive the oxygen abundance as
\begin{equation}
\frac{\rm O}{\rm H} = \frac{\rm O^+}{\rm H^+} + \frac{\rm O^{2+}}{\rm H^+}.
\label{o}
\end{equation}
The small fraction of the unseen O$^{3+}$ ion in the high-excitation
H {\sc ii} regions can be added to the oxygen abundance if the
He {\sc ii} $\lambda$4686 emission line is detected in their spectra.
Our sequences of photoionization models show that
O$^{3+}$/O is $\ga$1\% only in the highest-excitation H {\sc ii} regions with
O$^+$/(O$^+$+O$^{2+}$) $\la$ 0.1. In those H {\sc ii} regions,
O$^{3+}$/H$^{+}$ can be approximated by
\begin{equation}
\frac{\rm O^{3+}}{\rm H^+} = 0.5\times \frac{\rm He^{2+}}{\rm He^++He^{2+}}
\left(\frac{\rm O^{+}}{\rm H^+} + \frac{\rm O^{2+}}{\rm H^+}\right).
\label{o3p}
\end{equation}
This results in a minimal correction to the oxygen abundance derived just from
O$^+$ and O$^{2+}$. In the lower-excitation H {\sc ii} regions
with O$^+$/(O$^+$+O$^{2+}$) $>$ 0.1, we set O$^{3+}$/H$^{+}$ to zero.

The ionization structure of an H {\sc ii} region is determined by the balance
between the processes of photoionization and recombination. The rates
for some of these processes are poorly known. One of the major problems is
the uncertain rate of the dielectronic recombination for sulfur, chlorine
and argon ions. The comparison of the abundances of these elements
calculated with different assumed dielectronic recombination rates allows us
to put some constraints on these rates.
Fig. \ref{Fig5} shows the dependence of
$ICF$(S$^+$+S$^{2+}$) with  O$^+$/(O$^+$+O$^{2+}$)
%(***better avoid using
%$x$(O$^+$) = O$^+$/O since it may be ambiguous and does not correspond to
%the usual definition of x correct this in the figures too)
for the ``low $Z$'' models.
% with 12 +log O/H $\leq$ 7.6.
The solid
curve corresponds to what we call the ``nominal'' dielectronic recombination
rates (equal to the dielectronic recombination rates for second-row elements
in the same isoelectronic sequence). We adopt as the nominal dielectronic
recombination rates for S ions the same as for O ions. The dotted line
corresponds to dielectronic
recombination rates enhanced by a factor 3, and the dashed one is obtained
with a total suppression of
%low $T_e$
dielectronic recombination.  It is seen from
this figure  that $ICF$(S$^+$+S$^{2+}$) decreases when the
dielectronic recombination rate increases and the effect is more important for
high-excitation models. Figure \ref{Fig6} shows the values of S/O calculated
with different dielectronic recombination rates as a function of the
fraction O$^+$/(O$^+$+O$^{2+}$) for the galaxies in the
SDSS and HeBCD combined sample. Since the effect of dielectronic
recombination may depend on the metallicity in the H {\sc ii} region, via the
electron temperature,  we split the sample  into three
metallicity bins: the hottest H {\sc ii} regions with
12 + log O/H $\leq$ 7.6 (left panel),
the intermediate-temperature H {\sc ii} regions with 7.6 $<$ 
12 + log O/H $\leq$ 8.2 (middle panel) and
the coolest H {\sc ii} regions with 12 + log O/H $>$ 8.2 (right panel). 
It is expected that if the adopted rate of
dielectronic recombination for sulfur ions is correct then there should be
no dependence of the derived S/O with O$^+$/(O$^+$+O$^{2+}$). Inspection of
Fig. \ref{Fig6} shows that
%no
trends in the S/O abundance ratio are minimal if
a zero rate of dielectronic recombination for sulfur ions is adopted
for the low- and high-metallicity bins and the nominal rate is adopted for
the intermediate-metallicity bin. Thus,
we adopt the $ICF$s(S$^+$+S$^{2+}$) calculated with a zero
rate for dielectronic recombination for the 
low- and high-metallicity bins
and with a nominal rate for the intermediate-metallicity bin.

\begin{table*}
\caption{Element abundances in the SDSS galaxies$^1$ \label{tab2}}
\begin{tabular}{lccccccc} \hline
Spectrum number    &$t_e$(O {\sc iii})&12+log 
O/H&logN/O&logNe/O&logS/O&logAr/O&logFe/O \\
\hline
0266-51630-100& 1.5533$\pm$0.0327&7.84$\pm$0.02& --1.62$\pm$0.03& --0.85$\pm$0.04& --1.68$\pm$0.04& --2.40$\pm$0.04& --1.93$\pm$0.14 \\
0267-51608-384& 1.2055$\pm$0.0600&7.99$\pm$0.05& --1.30$\pm$0.07& --0.55$\pm$0.10& --1.71$\pm$0.11& --2.43$\pm$0.07& --1.69$\pm$0.21  \\
0270-51909-306& 0.9444$\pm$0.1699&8.33$\pm$0.20& --1.09$\pm$0.26& --0.77$\pm$0.41& --1.67$\pm$0.33&     ...      &     ...       \\
0272-51941-365& 1.0133$\pm$0.1661&8.38$\pm$0.18& --1.40$\pm$0.23& --0.65$\pm$0.36& --1.60$\pm$0.29& --2.50$\pm$0.22& --2.32$\pm$0.80 \\
0275-51910-429& 1.5117$\pm$0.0513&7.82$\pm$0.03& --1.45$\pm$0.05& --0.72$\pm$0.06& --1.73$\pm$0.10& --2.37$\pm$0.05&     ... \\ 
\hline
%0269-356& 7.97$\pm$0.10& --1.07$\pm$0.13& --0.85$\pm$0.23& --1.50$\pm$0.18& --2.19$\pm$0.12&    ...         & --1.69$\pm$0.19& --1.71$\pm$0.18 \\
%0270-306& 8.37$\pm$0.08& --1.12$\pm$0.08& --0.82$\pm$0.14& --1.72$\pm$0.21&    ...         &    ...         &    ...         & --1.90$\pm$0.17 \\ \hline
\end{tabular}

$^1$Only the first five galaxies are shown. The total list of galaxies
is available in electronic form. The Cl/O abundance ratios are not shown here 
for lack of space, but they are present in the electronic table.
\end{table*}

In a similar way, we consider the effect of dielectronic recombination on the
$ICF$s
for argon. We adopt as the nominal dielectronic recombination rate the same
values as for ions of neon. Figure \ref{Fig7} shows the dependence
of the Ar/O abundance ratio with O$^+$/(O$^+$+O$^{2+}$). Similarly to S/O, the
value of Ar/O should not depend on  O$^+$/(O$^+$+O$^{2+}$). A minimal
trend in the Ar/O
abundance ratio is seen for the H {\sc ii} regions in the whole
range of metallicities if the nominal rate is adopted.
A similar consideration for chlorine (Fig. \ref{Fig8}) shows that minimal
trends
are present for the H {\sc ii} regions with 12 + log O/H $>$ 7.6 if a
zero rate
is adopted. As for lower-metallicity H {\sc ii} regions, the statistics are
too poor to make definite conclusions.
We adopt above as the nominal dielectronic
recombination rates for Cl ions the same as for O ions.
%Thus we obtain that dielectronic recombination rates are small for
%sulfur and chlorine ions, but probably important for argon ions
%in the H {\sc ii} regions with 12 + log O/H $>$ 7.6.
In the following, we adopt the
$ICF$s derived from the models with a
nominal rate for dielectronic recombination
for argon ions and a zero rate for chlorine ions.

The adopted $ICF$s to obtain the elemental abundances
from the ionic ones are shown in Fig. \ref{Fig9}, as a function of
O$^+$/(O$^+$+O$^{2+}$) or O$^{2+}$/(O$^+$+O$^{2+}$). As in Fig. \ref{Fig3},
filled circles, open circles and stars correspond to the H {\sc ii} region
models with $Z$ = 0.02 $Z_\odot$, 0.05 $Z_\odot$ and 0.2 $Z_\odot$,
respectively. Dotted lines are
approximations of $ICF$s from the \citet{S90} models.

It is seen from Fig. \ref{Fig9} that there is very little
dependence of $ICF$(N$^+$),
$ICF$(Cl$^{2+}$), $ICF$(Ar$^{2+}$)
%$ICF$(Ar$^{2+}$+Ar$^{3+}$)
and $ICF$(Fe$^{2+}$)
on metallicity over the whole range of oxygen abundance considered here.
On the other hand,  $ICF$(Ne$^{2+}$) and $ICF$(S$^{+}$+S$^{2+}$)
are different for the
$Z$ = 0.02 and 0.05 $Z_\odot$ model sequences as compared to the
$Z$ = 0.2 $Z_\odot$ sequence.
This is due to the fact that the   $Z$ = 0.02
and 0.05 $Z_\odot$ model sequences include an X-ray ionizing source in
addition to the ionizing star population. This was required in order to
account for the large proportion of galaxies with nebular
He {\sc ii} $\lambda$4686. If an X-ray source is present, it creates in the
H {\sc ii} region an extended warm zone where hydrogen is partly recombined,
and where O$^{2+}$ ions are decoupled from Ne$^{2+}$ ions  due to the
efficiency of the O$^{2+}$ + H$^{0}$ charge transfer.
At present, we have no real clue on the nature and strength of these X-ray
sources, so the $ICF$ for neon in low excitation
H {\sc ii} regions with 12 + log O/H $\leq$ 8.2 is uncertain.
Fortunately, such cases do not represent a large fraction of our sample,
as can be seen in Figs. \ref{Fig6} -- \ref{Fig8}.

The new $ICF$s for nitrogen and iron are undistinguishable from the ones
obtained from the \citet{S90} models, while for other elements there are
significant differences. We fit the new $ICF$s by the expressions
\begin{eqnarray}
ICF({\rm N^+})&=-0.825v+0.718+0.853/v, &  {\rm low\ }Z, \nonumber\\
               &=-0.809v+0.712+0.852/v, &  {\rm intermed.\ }Z, \nonumber\\
               &=-1.476v+1.752+0.688/v, &  {\rm high\ }Z, \label{eq:Nn}
\end{eqnarray}
\begin{eqnarray}
ICF({\rm Ne^{2+}})&=-0.385w+1.365+0.022/w,&  {\rm low\ }Z, \nonumber\\
                   &=-0.405w+1.382+0.021/w,&  {\rm intermed.\ }Z, \nonumber\\
                   &=-0.591w+0.927+0.546/w,&    {\rm high\ }Z, \label{eq:Nen}
\end{eqnarray}
\begin{eqnarray}
ICF({\rm S^{+}+S^{2+}})&=0.121v+0.511+0.161/v, \ {\rm low\ }Z, \nonumber\\
                     &=0.155v+0.849+0.062/v, \ {\rm inter.\ }Z, \nonumber\\
                        &=0.178v+0.610+0.153/v, \  {\rm high\ }Z, \nonumber\\
\label{eq:Sn}
\end{eqnarray}
\begin{eqnarray}
ICF({\rm Cl^{2+}})&=0.756v+0.648+0.128/v, \ {\rm low\ }Z, \nonumber\\
                &=0.814v+0.620+0.131/v, \ {\rm intermed.\ }Z, \nonumber\\
                   &=1.186v+0.357+0.121/v, \  {\rm high\ }Z, \label{eq:Cl2n}
\end{eqnarray}
\begin{eqnarray}
ICF({\rm Ar^{2+}})&=0.278v+0.836+0.051/v, \ {\rm low\ }Z, \nonumber\\
                &=0.285v+0.833+0.051/v, \ {\rm intermed.\ }Z, \nonumber\\
                   &=0.517v+0.763+0.042/v, \ {\rm high\ }Z,
\label{eq:Ar1n}
\end{eqnarray}
\begin{eqnarray}
ICF({\rm Ar^{2+}+Ar^{3+}})&=0.158v+0.958+0.004/v, \ {\rm low\ }Z, \nonumber\\
                       &=0.104v+0.980+0.001/v, \ {\rm int.\, }Z, 
\nonumber\\
                           &=0.238v+0.931+0.004/v,  {\rm high\ }Z, \nonumber\\
\label{eq:Ar2n}
\end{eqnarray}
\begin{eqnarray}
ICF({\rm Fe^{2+}})&=0.036v-0.146+1.386/v, &  {\rm low\ }Z, \nonumber\\
                 &=0.301v-0.259+1.367/v, &  {\rm intermed.\ }Z, \nonumber\\
                 &=-1.377v+1.606+1.045/v, &  {\rm high\ }Z, \label{eq:Fen}
\end{eqnarray}
where $v$=O$^+$/(O$^+$+O$^{2+}$) and $w$=O$^{2+}$/(O$^+$+O$^{2+}$).
The fits given by Eqs. \ref{eq:Nn}, \ref{eq:Nen} and \ref{eq:Fen} are
applicable over the whole range of $v,w$ = 0 -- 1,
while the remaining fits are applicable
only for $v$ $\la$ 0.8. We note however that only a few H {\sc ii} regions
in our sample have $v$ $>$ 0.8.
  We adopt the ``low $Z$'' $ICF$s for
H {\sc ii} regions with 12+log O/H $\leq$ 7.2, ``high $Z$'' $ICF$s
for H {\sc ii} regions with 12+log O/H $\geq$ 8.2 and linearly
interpolated $ICF$s between ``low $Z$'' and ``intermediate $Z$'' cases
for H {\sc ii} regions with
12+log O/H between 7.2 and 7.6, and between ``intermediate $Z$'' and
``high $Z$'' cases for H {\sc ii} regions with 12+log O/H between 7.6 and 8.2.
Although $ICF$s for some elements (e.g., for nitrogen) appear very similar
in all metallicity bins (Fig. \ref{Fig9}), we show in Eqs. \ref{eq:Nn} --
\ref{eq:Fen} fits obtained separately for each bin.

In such a way, we can derive all elemental abundances using
only analytical formulae. This is useful for the derivation of the formal
uncertainties in the derived abundances, since they can be obtained by
expanding the formulae in Taylor series, and do not require the use of
Monte-Carlo methods.

\section{Discussion \label{sect4}}% of the observed abundance patterns}

\subsection{SDSS DR3 galaxies versus galaxies from the HeBCD sample}

\begin{figure*}[t]
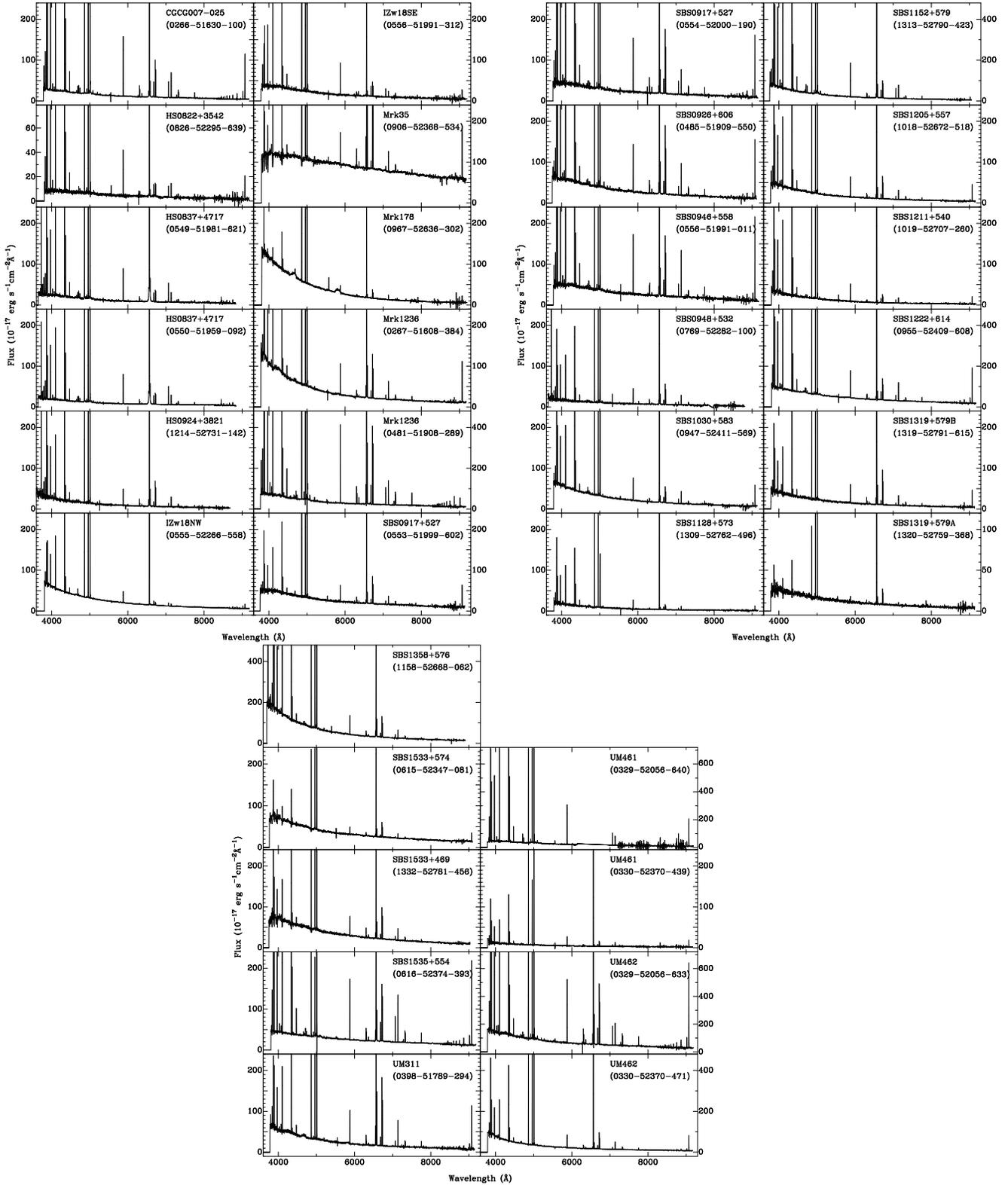

\hspace*{-0.0cm}\psfig{figure=3763f10a.ps,angle=0,width=8.5cm}%,clip=}
\hspace*{0.5cm}\psfig{figure=3763f10b.ps,angle=0,width=8.5cm,clip=}
\hspace*{4.0cm}\psfig{figure=3763f10c.ps,angle=0,width=8.5cm}%,clip=}
%\hspace*{0.3cm}\psfig{figure=nnesarDR3new3_1.ps,angle=0,height=10.5cm}%,clip=}
%\hspace*{0.3cm}\psfig{figure=nnesarDR3new3_2.ps,angle=0,height=10.5cm,clip=}
\caption{SDSS spectra of the H {\sc ii} regions listed in Table \ref{tab3}.
The names of the spectra in the SDSS database are shown in parentheses.}
\label{Fig10}
\end{figure*}

Electron temperatures $t_e$(O {\sc iii}), element abundances and
abundance ratios in the H {\sc ii} regions from the SDSS
DR3 sample calculated as explained above are shown in Table
\ref{tab2}. Some objects from the SDSS sample are also present in the
HeBCD sample. For those objects we compare in Table \ref{tab3} oxygen
abundances derived from different observations.
The SDSS spectra of these galaxies are shown in Fig. \ref{Fig10}.
Since the redshifts of these
galaxies are small, the [O {\sc ii}] $\lambda$3727 is not in the 
observed wavelength range of the SDSS spectra and
therefore the O$^+$ abundance was obtained from the
[O {\sc ii}] $\lambda$7320, 7331 emission line fluxes.
Exceptions are HS 0837+4717, HS 0924+3821,
SBS 0948+532 and SBS 1358+576 with higher redshifts where the O$^+$ abundance
was obtained from the
[O {\sc ii}] $\lambda$3727 emission line fluxes.
The galaxies HS 0837+4717, Mrk 1236, SBS 0917+527,
UM 461 and UM 462 were observed twice
in the course of SDSS and we show both oxygen abundances for each of those
galaxies. It is seen from
Table \ref{tab3} that 12+log O/H derived from different observations are
generally
in good agreement. One exception is the galaxy UM 461. In one of two SDSS
observations
of this galaxy, the [O {\sc iii}] $\lambda$4959, 5007 emission lines are
clipped resulting in a low oxygen abundance. The oxygen abundances for some
galaxies from Table \ref{tab3} are also in agreement with independent
abundance determinations from the same SDSS spectra by \citet{K04a,K04b}.

\begin{figure*}[t]
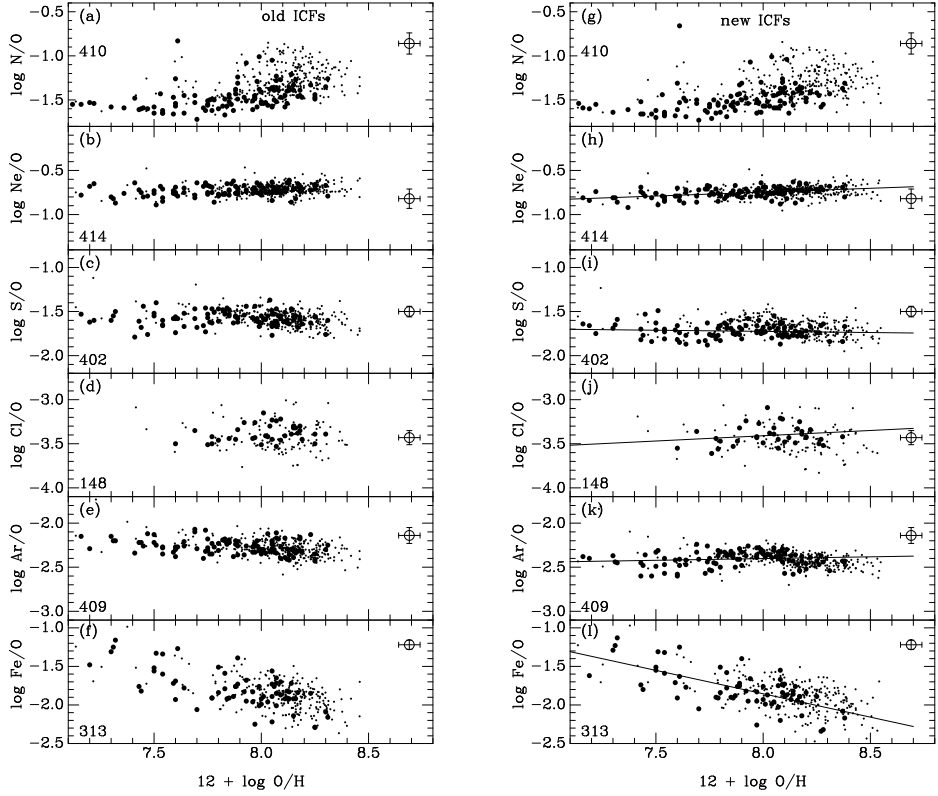

\hspace*{2.0cm}\psfig{figure=3763f11a.ps,angle=0,height=10.5cm}%,clip=}
%\hspace*{0.3cm}\psfig{figure=nnesarDR3new3_1.ps,angle=0,height=10.5cm}%,clip=}
\hspace*{1.0cm}\psfig{figure=3763f11b.ps,angle=0,height=10.5cm}%,clip=}
\caption{log N/O (a, g), log Ne/O (b, h), log S/O (c, i),
log Cl/O (d, j), log Ar/O (e, k) and log Fe/O (f, l) vs oxygen
abundance 12 + log O/H for the emission-line galaxies. Large filled
circles are galaxies from the HeBCD sample, dots are galaxies
from the SDSS sample.
Abundances in the left panel are calculated with the expressions used by
\citet{ITL94,ITL97} and \citet{TIL95} and those in
the right panel are obtained with expressions from this paper. 
The solar values as compiled by \citet{L03} are indicated by the large open
circles and the associated error bars are shown. Straight lines are
linear regressions obtained for the HeBCD sample.
}
\label{Fig11}
\end{figure*}

In Fig. \ref{Fig11} we show the computed N/O, Ne/O, S/O, Cl/O,
Ar/O, Fe/O abundance ratios
vs oxygen abundance 12 + log O/H for the SDSS and HeBCD merged sample
using old (left panels) and new (right panels) expressions to derive the
abundances.
In all panels, the HeBCD galaxies are represented by large filled
circles, while the SDSS galaxies are shown by dots. In each panel,
the solar abundance ratio \citep[as compiled by ][]{L03} is represented
by a large open circle.

\begin{table}
\caption{Comparison of element abundances in the galaxies present in both
HeBCD and SDSS samples \label{tab3}}
\begin{tabular}{lcc} \hline
Name    &(12+logO/H)$_{\rm SDSS}$&(12+logO/H)$_{\rm HeBCD}$ \\
\hline
CGCG007--025       & 7.84 $\pm$ 0.02        & 7.78 $\pm$ 0.01 \\ %  0266-093 1
HS0822+3542        & 7.42 $\pm$ 0.03        & 7.45 $\pm$ 0.01 \\ %2 0826-639 2
HS0837+4717        & 7.63 $\pm$ 0.03        & 7.61 $\pm$ 0.01 \\ %  0549-621 1
                    & 7.60 $\pm$ 0.03        &                 \\ %  0550-092 1
HS0924+3821        & 8.15 $\pm$ 0.04        & 8.09 $\pm$ 0.02 \\ %  1214-142 3
IZw18NW            & 7.14 $\pm$ 0.05        & 7.16 $\pm$ 0.02 \\ %2 0555-558 2
IZw18SE            & 7.25 $\pm$ 0.04        & 7.19 $\pm$ 0.03 \\ %  0556-312 1
Mrk35              & 8.29 $\pm$ 0.12        & 8.37 $\pm$ 0.01 \\ %  0906-534 2
Mrk178             & 7.90 $\pm$ 0.11        & 7.82 $\pm$ 0.02 \\ %  0967-302 3
Mrk1236            & 7.99 $\pm$ 0.05        & 8.09 $\pm$ 0.03 \\ %  0267-384 1
                    & 8.12 $\pm$ 0.02        &                 \\ %  0481-289 1
SBS0917+527        & 7.81 $\pm$ 0.04        & 7.87 $\pm$ 0.02 \\ %  0553-602 1
                    & 7.91 $\pm$ 0.02        &                 \\ %  0554-190 1
SBS0926+606        & 7.99 $\pm$ 0.02        & 7.92 $\pm$ 0.02 \\ %  0485-550 1
SBS0946+558        & 8.04 $\pm$ 0.02        & 8.01 $\pm$ 0.01 \\ %  0556-011 1
SBS0948+532        & 8.04 $\pm$ 0.03        & 8.08 $\pm$ 0.01 \\ %2 0769-100 2
SBS1030+583        & 7.79 $\pm$ 0.03        & 7.80 $\pm$ 0.01 \\ %2 0947-569 2
SBS1128+573        & 7.72 $\pm$ 0.05        & 7.75 $\pm$ 0.03 \\ %  1309-496 3
%SBS1132+503        & 8.10 $\pm$ 0.07        & 7.87 $\pm$ 0.15 \\
SBS1152+579        & 7.95 $\pm$ 0.02        & 7.88 $\pm$ 0.01 \\ %  1313-423 3
SBS1205+557        & 7.92 $\pm$ 0.04        & 7.77 $\pm$ 0.03 \\ %  1018-518 3
SBS1211+540        & 7.65 $\pm$ 0.03        & 7.64 $\pm$ 0.01 \\ %  1019-260 3
SBS1222+614        & 8.00 $\pm$ 0.02        & 7.97 $\pm$ 0.01 \\ %2 0955-608 2
SBS1319+579B       & 8.40 $\pm$ 0.10        & 8.10 $\pm$ 0.01 \\ %  1319-615 3
SBS1319+579A       & 7.75 $\pm$ 0.18        & 8.17 $\pm$ 0.04 \\ %  1320-368 3
SBS1358+576        & 7.94 $\pm$ 0.03        & 7.89 $\pm$ 0.02 \\ %  1158-062 3
SBS1533+574        & 8.11 $\pm$ 0.12        & 8.14 $\pm$ 0.02 \\ %2 0615-081 2
SBS1533+469        & 8.16 $\pm$ 0.06        & 7.99 $\pm$ 0.02 \\ %  1332-456 3
SBS1535+554        & 8.08 $\pm$ 0.02        & 8.08 $\pm$ 0.04 \\ %2 0616-393 2
UM311              & 8.40 $\pm$ 0.11        & 8.38 $\pm$ 0.04 \\ %  0398-294 1
UM461              & 7.81 $\pm$ 0.02        & 7.79 $\pm$ 0.03 \\ %  0329-640 1
                    & ~\,7.28 $\pm$ 0.04$^1$ &                 \\ %2 0330-439 2
UM462              & 7.80 $\pm$ 0.02        & 7.90 $\pm$ 0.02 \\ %  0329-633 1
                    & 7.87 $\pm$ 0.03        &                 \\ 
\hline %2 0330-471 2
\end{tabular}

$^1$ [O {\sc iii}] $\lambda$4959, 5007 emission lines are clipped.
\end{table}

It is seen from Fig. \ref{Fig11} that SDSS
galaxies are in general more metal-rich than the galaxies from
the HeBCD sample. Only a few SDSS H {\sc ii} regions have
an oxygen
abundance 12+log O/H smaller than 7.6. Two of them are actually
the northwest and the southeast
components of the well-known metal-deficient star-forming dwarf galaxy
I Zw 18, which are also present in the HeBCD sample.
Thus,  star-forming galaxies
with extremely low metallicity are very rare in the spectroscopic data base of
the SDSS.

%%%%%%%%%%%%%%%%%%%%%%%%%%%%%%%%%%%%%%%%%%%%%%%%
%    Fig.10
%%%%%%%%%%%%%%%%%%%%%%%%%%%%%%%%%%%%%%%%%%%%%%%%
%%%%%%%%%%%%%%%%%%%%%%%%%%%%%%%%%%%%%%%%%%%%%%%%%

%%%%%%%%%%%%%%%%%%%%%%%%%%%%%%%%%%%%%%%%%%%%%%%%
%    Fig.11
%%%%%%%%%%%%%%%%%%%%%%%%%%%%%%%%%%%%%%%%%%%%%%%%
%%%%%%%%%%%%%%%%%%%%%%%%%%%%%%%%%%%%%%%%%%%%%%%%%

%%%%%%%%%%%%%%%%%%%%%%%%%%%%%%%%%%%%%%%%%%%%%%%%
%    Fig.12
%%%%%%%%%%%%%%%%%%%%%%%%%%%%%%%%%%%%%%%%%%%%%%%%
\begin{figure*}[t]
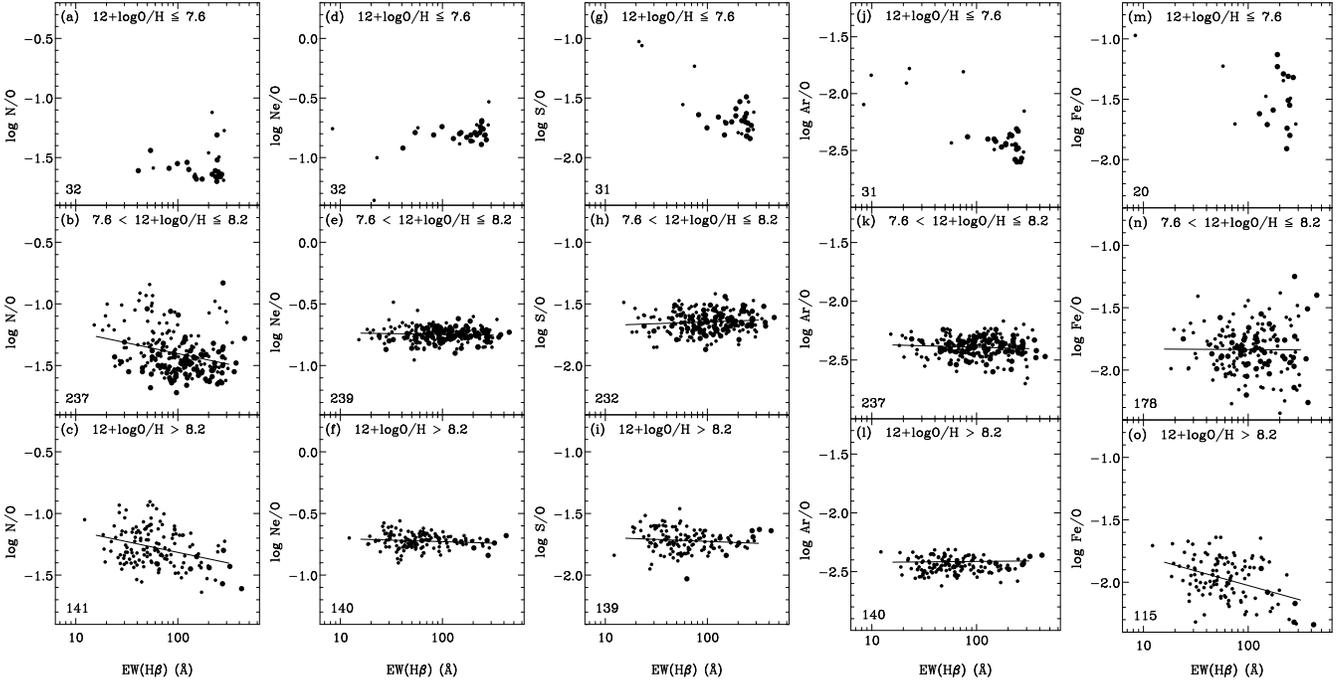

\hspace*{-0.0cm}\psfig{figure=3763f12a.ps,angle=0,height=9.0cm}%,clip=}
\hspace*{0.1cm}\psfig{figure=3763f12b.ps,angle=0,height=9.0cm}%,clip=}
\hspace*{0.1cm}\psfig{figure=3763f12c.ps,angle=0,height=9.0cm}%,clip=}
\hspace*{0.1cm}\psfig{figure=3763f12d.ps,angle=0,height=9.0cm}%,clip=}
\hspace*{0.1cm}\psfig{figure=3763f12e.ps,angle=0,height=9.0cm}%,clip=}
\caption{{\bf (a)--(c)} Nitrogen-to-oxygen abundance ratio, log N/O,
vs equivalent width
of H$\beta$ emission line, EW(H$\beta$), for the galaxies in three
metallicity bins.
{\bf (d)--(f)} The same as in {\bf (a)--(c)}, but for neon-to-oxygen
abundance ratio.
{\bf (g)--(i)} The same as in {\bf (a)--(c)}, but for sulfur-to-oxygen
abundance ratio.
{\bf (j)--(l)} The same as in {\bf (a)--(c)}, but for argon-to-oxygen
abundance ratio.
{\bf (m)--(o)} The same as in {\bf (a)--(c)}, but for iron-to-oxygen
abundance ratio.
%Galaxies with [O {\sc iii}] $\lambda$4959/H$\beta$ $<$ 0.7 are not shown.
Straight lines are linear regressions obtained for intermediate $Z$ and high
$Z$ samples.
}
\label{Fig12}
\end{figure*}
%%%%%%%%%%%%%%%%%%%%%%%%%%%%%%%%%%%%%%%%%%%%%%%%%

The distribution of the SDSS DR3 galaxies
in the different panels of Fig. \ref{Fig11} confirms previous
findings concerning HeBCD \citep{TIL95,IT99}
and SDSS Early Data Release galaxies \citep{I04}.
The use of the new expressions to compute elemental abundances
(Fig.\ref{Fig11},
middle and right panels) result in very slight changes.
Ne/O abundance ratios are typically decreased
by 0.1 dex. The largest
effect is for the S/O and Ar/O abundance ratios, which decrease typically
by 0.15 dex.

The mean Ne/O and Cl/O abundance ratios for our galaxies are in good agreement
with the solar abundance ratios from
\citet{L03} (open circles in Fig. \ref{Fig11}). On the other hand,
the S/O and Ar/O abundance ratios in the H {\sc ii} regions are lower than
the solar abundance ratios (by 0.15 dex and 0.25 dex, respectively). We cannot
exclude the possibility that these
differences are caused by uncertainties in the determination of S and Ar
abundances in the H {\sc ii} regions. Alternatively, they may indicate that
the solar S and Ar abundances are not as accurately known as thought.

  We show by solid lines in Fig. \ref{Fig11} (middle and right panels) the
regression lines for log Ne/O, log S/O, log Cl/O, log Ar/O and log Fe/O vs
12+log O/H obtained from the HeBCD sample only. The equations of these
lines are the following:
\begin{equation}
\log \frac{\rm Ne}{\rm O} = (0.088\pm0.018){\rm X} - (1.450\pm0.144),
\label{regrNe}
\end{equation}
\begin{equation}
\log \frac{\rm S}{\rm O} = -(0.026\pm0.027){\rm X} - (1.514\pm0.215),
\label{regrS}
\end{equation}
\begin{equation}
\log \frac{\rm Cl}{\rm O} = (0.120\pm0.106){\rm X} - (4.365\pm0.852),
\label{regrCl}
\end{equation}
\begin{equation}
\log \frac{\rm Ar}{\rm O} = (0.039\pm0.031){\rm X} - (2.714\pm0.244),
\label{regrAr}
\end{equation}
\begin{equation}
\log \frac{\rm Fe}{\rm O} = -(0.606\pm0.081){\rm X} + (2.994\pm0.639),
\label{regrFe}
\end{equation}
where X = 12 + log O/H.

For the merged HeBCD and SDSS sample we  find no significant trends with the 
oxygen
abundance for the S/O, Cl/O and Ar/O abundance ratios in the whole range
of oxygen abundance (Figs. \ref{Fig11}h -- \ref{Fig11}k).
The best derived abundances among the $\alpha$-elements are
for oxygen and neon because the atomic data for these elements are better
known, the abundances are derived from bright emission lines and only two
ionization stages are present in significant amount in  H {\sc ii} regions.
%The precision of S, Cl and Ar abundance determination is not as good.

\subsection{Elemental depletion onto dust grains}

We note that there is a slight increase in Ne/O with increasing O/H in panel h
of Fig. \ref{Fig11} (the correlation is significant at the 99.98\% level
according to the Spearman statistical test). If the new expressions for
deriving abundances are correct then this slight  increase (by
$\sim$ 0.1 dex over the entire metallicity range considered) can be
interpreted by depletion of oxygen
onto dust grains. This corresponds to
$\sim$ 20\% of oxygen locked in the dust grains in the highest-metallicity
H {\sc ii} regions of our sample, while no significant depletion would be
present in the H {\sc ii} regions with lower metallicity (assuming that the
adopted solar Ne/O ratio corresponds to that of the interstellar medium).
Since Ar is a noble gas like Ne, one would expect the same trend in Ar/O as in
Ne/O. The regression lines for the merged
HeBCD and SDSS sample shown
in Fig. \ref{Fig11}h and \ref{Fig11}k are
\begin{equation}
\log \frac{\rm Ne}{\rm O} = (0.097\pm0.015){\rm X} - (1.531\pm0.120),
\label{regrNe1}
\end{equation}
\begin{equation}
\log \frac{\rm Ar}{\rm O} = -(0.084\pm0.021){\rm X} - (1.729\pm0.166),
\label{regrAr1}
\end{equation}
It is seen that the slopes of the Ne/O vs O/H and Ar/O vs O/H
regressions are statistically different. This could
indicate that the production of Ar relative to O by the
stellar populations decreases with increasing metallicity.

As seen in  Fig. \ref{Fig11}, iron has a very different behavior as compared
with other elements. The Fe/O ratio is well below solar for most of the
galaxies. In addition, the trend with O/H is opposite to that in metal-poor
stars in our Galaxy \citep{C00,T03,B04}. This  argues in
favor  of depletion of iron onto dust grains. The decrease of the Fe/O
abundance ratio with increasing O/H (by $\sim$ 1 dex when
the oxygen abundance increases by $\sim$ 1.5 dex
as seen in Fig. \ref{Fig11}l) implies that  depletion of Fe increases
with increasing metallicity, similar to what is suggested by
\citet{R05} from the consideration of a small
sample of Galactic and extragalactic H {\sc ii} regions. If this explanation
is correct then only
$\la$ 10\% of iron in the
high-metallicity H {\sc ii} regions of our sample (log O/H +12 = 8.2 -- 8.5)
is in gas phase and the remaining
$\ga$ 90\% of iron is locked in dust grains, while in the H {\sc ii}
regions with the lowest metallicity the depletion is much lower. This
conclusion is supported by the small trend seen in the Ne/O abundance ratio.
Since Ne is not locked in dust grains, the increase of Ne/O with
metallicity can be explained by a small depletion of oxygen.
Since oxygen is an element $\sim$ 15 times more abundant than iron,
a depletion of oxygen by $\sim$ 20\% would correspond to an important iron
depletion. In the panels on the right of  Fig. \ref{Fig12}, we show the
variation of
Fe/O as a function of the equivalent width of H$\beta$, 
EW(H$\beta$), for the galaxies in the SDSS and HeBCD merged sample.
Fig. \ref{Fig12}m corresponds to the lowest metallicity bin, defined by
12 + log O/H $\leq$ 7.6, Fig. \ref{Fig12}n corresponds to the intermediate
metallicity bin (7.6 $<$ 12 + log O/H $\leq$ 8.2) and
Fig. \ref{Fig12}o corresponds
to the highest metallicity bin in our sample (12 + log O/H $>$ 8.2). If we
take EW(H$\beta$)
as a measure of the age of the ionizing stars \citep[but see][]{SI03},
Fig. \ref{Fig12}o strongly suggests that the grains are
gradually destroyed on a time scale of several Myr. The Spearman rank
correlation coefficient is $R_S$= --0.30 with an associated
%probability of 1.2$\times$10$^{-3}$,
significance level of 99.9\%,
indicating that the observed correlation is not due to chance. There is
no similar clear trend in Fig. \ref{Fig12}n ($R_S$= --0.006 with an associated
%probability of 0.93).
significance level of 7\%).
However, since the initial depletion
in the intermediate metallicity bin is lower than in the highest metallicity
bin, detecting a temporal variation of the correlation is more difficult. Note
that the scatter in Fe/H at any value of EW(H$\beta$) as well as at any
value of O/H is far more important in Fe/O than in Ne/O, S/O, Cl/O and Ar/O.
We believe that this is mostly due  to errors in the iron abundance
determinations since the [Fe {\sc iii}] lines are generally very weak but one
cannot exclude  that the degree of iron depletion might vary among H {\sc ii}
regions of similar characteristics.

\subsection{The nitrogen conundrum}

Much has already been written about the N/O versus O/H diagram (upper line of
panels in Fig. \ref{Fig11}) \citep[e.g., ][]{G90,P92,H00}.
Our new results basically confirm the findings
in our previous papers, based on considerably
smaller samples \citep{IT99,I04}:
the N/O ratio decreases as O/H decreases but reaches a
plateau at 12 + log O/H $\leq$ 7.6.
Moreover, while at a given O/H, there is a
significant dispersion in N/O when 12 + log O/H $>$ 7.6, the dispersion is
very small at lower values of O/H. The lack of H {\sc ii} regions with
12 + log O/H $\leq$ 7.6 and high N/O is likely not due to selection
effects. Indeed higher N/O abundance ratio implies higher flux
of the [N {\sc ii}] $\lambda$6584 emission line which can be more easily
detected in this case. On the other hand, it is unlikely that we missed
low-metallicity galaxies with a very low N/O abundance ratio, below the
plateau value of log N/O $\sim$ --1.6, as the [N {\sc ii}] $\lambda$6584
emission line was detected in all most metal-deficient galaxies.

The N/O versus O/H pattern for H {\sc ii} regions differs significantly from
that for damped Ly$\alpha$ systems (DLA) \citep[e.g.,][]{C03,M05},
in that some  DLAs have N/O ratios much lower than --1.6 dex, the value of the
observed plateau for H {\sc ii} regions.
Recently, the N/O -- O/H diagram has also been
constructed for metal-poor stars \citep{Sp05}. Stars affected by mixing
have been excluded from their diagram. It is seen that some very
low-metallicity stars also have values of N/O below the
H {\sc ii} region plateau, although they
are larger than the lowest N/O found for DLAs. To interpret these results
in terms of models of chemical evolution of galaxies, we
should however bear in mind that
abundances determined in DLAs \citep[see][]{D04} as well as in stars
\citep{Sp05} are generally less accurate than the N/O ratios in metal-poor
H {\sc ii} regions where the main source of uncertainty is the measurement of
the weak [N {\sc ii}] 6584 line. In the HeBCD sample, the average formal
error in N/O is $\la$ 0.1 dex at 12 + log O/H $\leq$ 7.7, and it
is $\leq$ 0.2 dex for bright galaxies in the SDSS DR3 sample at the same
metallicities. Also, DLAs, Galactic halo stars and metal-poor emission-line
galaxies had likely very different star formation histories. The differences
in the abundance patterns observed between these systems might also result
from the different timescales and intensities of the star formation episodes
which enriched their interstellar medium.

The fact that no H {\sc ii} region is found  with an extremely low N/O
abundance ratio ( log N/O $\leq$ --1.6)  suggests that the SDSS 
galaxies are not
extremely young and have ages more than 100 -- 300 Myr, which are required
for the completion of the evolution of the intermediate-mass stars.

We also confirm the trend of increasing N/O with decreasing EW(H$\beta$)
already mentioned by \citet{I04}.  This trend is seen in the
12 + log O/H $>$ 8.2 metallicity bin (at a
%1.1$\times$10$^{-6}$
99.99\%
significance level according to
the Spearman statistics) and also in the 7.6 $<$ 12 + log O/H $<$ 8.2
metallicity bin (at a
%1.4$\times$10$^{-3}$
99.8\% significance level)
(panels c and b of
Fig. \ref{Fig12}). Note that, apart from Fe/O, which we discussed before, none
of the remaining abundance ratios shown in Fig. \ref{Fig12} exhibits such a
trend. Therefore, the trend for N/O with EW(H$\beta$) must be real and not due
to selection effects or systematic errors in abundance determinations.

Since EW(H$\beta$) is a measure of the age of the ionizing-star population,
it is interesting to check whether the larger
N/O abundance ratio in the more evolved starbursts is due to the nitrogen
enrichment by the Wolf-Rayet stars.

Let us consider the massive star rotating models for a metallicity equal
to $Z$=0.004, which corresponds to 12+log O/H=8.2, computed recently
by \citet{MM05}.
The 60$M_\odot$ star model loses about half of its initial mass
during the first 4.76 Myr. During this phase the amount of nitrogen in the
wind ejected material is about 9 times the initial content
(the initial content would be the mass of nitrogen ejected if no change of
nitrogen abundance occurred). On the other hand, oxygen appears to be depleted
by a factor 2.5 with respect to its initial value.
Thus,
we conclude that if one were to observe only wind material of this kind
then one would expect an increase of the N/O ratio by about a factor 20
in a few Myrs. A similar estimate made on the basis of a 120$M_\odot$
stellar model leads to an N/O increase by a factor 55, while a 30$M_\odot$
stellar model would predict an increase of only a factor 5.
Of course, there is a dilution of the WR material
with the local interstellar material of the H {\sc ii} region
\footnote{
Note also that these stars may also eject through their winds and at
the time of the supernova explosion material with very low N/O ratios.
This would also concur, together with the dilution effect, to
decrease the N/O ratio in the H {\sc ii} region.}.
The enhancement
factor of the N/O abundance ratio in the H {\sc ii} region due to the WR wind
can be estimated as

\begin{equation}
\frac{{\rm N/O} ({\rm HII})}{{\rm N/O} ({\rm ini})} =
\frac{p({\rm N})\times (M_{w}/M_*) \times (M_*/M({\rm HII}))+1}
{p({\rm O})\times (M_{w}/M_*) \times (M_*/M({\rm HII}))+1},
  \label{eq:Nenh}
\end{equation}
where $p$(N) and $p$(O) are the production factors of nitrogen and
oxygen in the wind, $M_{w}$ is the total mass in the winds, $M_*$ is the
integrated initial mass of the stars and $M$(H {\sc ii}) is the mass of the
H {\sc ii} region.
Let us take $p$(N)=9, $p$(O)=0.4, typical of the 60$M_\odot$ model. In the
mass range between 30 and 120 $M_\odot$, on average about half of the stellar
mass is lost through stellar winds enriched in N and depleted in O, thus
\begin{equation}
M_{w}/M_*\approx {\int_{30}^{120}
0.5M M^{-2.35} {\rm d}M \over \int_{0.1}^{120}M M^{-2.35}
{\rm d}M}\sim 0.03. \label{eq:Mw}
\end{equation}
About 3 percent of the mass in stars is ejected under the form of N-rich
stellar wind.
For an H {\sc ii} region of constant number density $n$, the value of
$M_*/M$(H {\sc ii}) for a cluster with a Salpeter initial mass function is
roughly equal to $n$/360. If the number
density of the H {\sc ii} region is 100 cm$^{-3}$, $M_*/M$(H {\sc ii})
is about 0.28. Thus the
enhancement factor of the N/O abundance ratio in the H {\sc ii} region is
$\sim$ 1.07 (or $\sim$ 0.03 dex), significantly lower than the
dispersion of the N/O abundance ratio in Fig. \ref{Fig11} and the value of
the trend seen in Fig. \ref{Fig12} (left panel).

It is likely, therefore,
that the large dispersion of N/O abundance ratio in Fig. \ref{Fig11} and
the trend seen in Fig. \ref{Fig12} are not due to a global increase of the
N/O ratio in the H {\sc ii} region, but it could still be  a local effect.
It was assumed
in the above consideration that the H {\sc ii} region is uniform, i.e. its
density is constant. However, it is very likely that the density of the N
enriched ejecta is much higher than that in the ambient H {\sc ii} region.
Since the luminosity of the forbidden lines in the low-density medium scales
as the number density of the emitting ions multiplied by the number density of
electrons, the apparent enhancement of the N/O abundance
ratio, can be estimated from
\begin{equation}
\left[\frac{{\rm N/O} ({\rm HII})}{{\rm N/O} ({\rm ini})}\right]_{\rm app} =
\frac{p({\rm N})\times n_{\rm ej}/n \times M_{w}/M({\rm HII})+1}
{p({\rm O})\times n_{\rm ej}/n \times M_{w}/M({\rm HII})+1}, \label{eq:Nenh1}
\end{equation}
where $n_{\rm ej}$ is {the number density of gas particles  in the
ejecta}. Adopting $n$ = 100 cm$^{-3}$ and
$n_{\rm ej}$ = 1000 cm$^{-3}$ we obtain an apparent
enhancement factor of $\sim$ 1.7 from Eq. \ref{eq:Nenh1}, comparable to the
dispersion of
the N/O abundance ratio in Fig. \ref{Fig11} and the value of the trend in
Fig. \ref{Fig12}. Note that there is no direct diagnostic of the density of
the emitting ejecta. The density derived from the [S {\sc ii}] line ratio
applies essentially to the H {\sc ii} since the winds are not enhanced in
sulfur.

In summary, we find that the observed trend of  N/O increasing as
EW(H$\beta$) decreases is naturally explained by the expected ejection from
Wolf-Rayet stars. However,  the true value of the
N/O enhancement factor is defined by Eq. \ref{eq:Nenh}, and is small.
The dispersion of the observed
N/O abundance ratio is caused by the local enrichment by the WR stars and
the value of the N/O abundance ratio in the  H {\sc ii} before enrichment by
the present starburst is likely defined by the lower boundary
in the N/O vs O/H diagram. If this is correct then the small dispersion of
the N/O abundance ratio in the most-metal deficient H {\sc ii} regions could
perhaps
be explained by the very low number of the WR stars and short WR stage during
the low-metallicity burst of star formation. However, a larger sample of
extremely low-metallicity objects would be needed to confirm this.

\section{Conclusions \label{sect5}}

We have examined all the galaxies in the Sloan Digital Sky Survey
(SDSS) Data Release 3 (DR3) to select out those
with a detected [O {\sc iii}] $\lambda$4363 emission line, which allows, in 
principle, a direct
element abundance determination based on the electron temperature.
To minimize errors in the abundance determinations, we have kept for 
further analysis only $\sim$ 310 SDSS DR3 galaxies that are sufficiently 
bright in H$\beta$ ($F$(H$\beta$) $>$ 10$^{-14}$ erg 
s$^{-1}$ cm$^{-2}$). The [O {\sc iii}] $\lambda$4363 line in these
objects is detected at the $\ga$ 2$\sigma$ level. This sample
was merged with a sample of 109 blue compact dwarf galaxies with 
high-quality observations and 
large equivalent widths of emission lines, 
referred to as the HeBCD sample. In this way we obtained samples of
emission-line galaxies, spanning a range in metallicities from
12 + log O/H  $\sim$ 7.1 ($Z_\odot$/30) to
$\sim$ 8.5 (0.7 $Z_\odot$). Our main conclusions are as follows:

1. Despite an examination of the entire SDSS DR3 sample of $\sim$ 530 000
galaxies, we found only 6 new
galaxies with extremely low metallicity (12 + log O/H $<$ 7.6, i.e. $Z$ $<$
$Z_\odot$/12), in addition to the galaxies discovered by \citet{K04a}.
For comparison, the HeBCD sample
which consists of only $\sim$100 galaxies contains about 15 such galaxies.
This difference is in part due to selection effects.
The selection criteria for the HeBCD sample favour extremely
low-metallicity galaxies as they were collected for the determination of
the primordial He abundance. The small
number of  extremely low-metallicity galaxies in the SDSS may suggest that
such galaxies are actually very rare. However, the selection effects in the
SDSS might be
important because most metal-poor galaxies are in general low-luminosity
objects and the SDSS spectroscopic data are only complete for galaxies
brighter than $\sim$ 17.7 mag \citep{I04}.

2. The emission-line galaxies in the SDSS sample show
distributions of the N/O, Ne/O, S/O, Cl/O, Ar/O, Fe/O abundance ratios vs
oxygen abundance similar to those found
earlier for the galaxies in the HeBCD sample
\citep{TIL95,IT99,G03}.

3. The $\alpha$ element-to-oxygen abundance ratios Ne/O, Cl/O, S/O and Ar/O
do not show large trends with oxygen abundance. The best determined ratio,
Ne/O, increases slightly  with increasing O/H (by $\sim$ 0.1 dex in the
considered metallicity range). This can be
explained by the depletion of oxygen onto dust grains with $\sim$ 20\% of O
locked in dust. No significant depletion of oxygen is found in the
lower-metallicity H {\sc ii} regions.
The mean Ne/O and Cl/O abundance ratios
in the H {\sc ii} regions from our sample are consistent with the solar ratios.
On the other hand, the mean S/O and Ar/O abundance ratios in the H {\sc ii}
regions are lower than the solar ratio. This may indicate the S and Ar
abundances in the Sun are somewhat uncertain.

4. The Fe/O abundance ratio shows 
%a significant 
an underabundance of iron
relative to oxygen as compared to solar, suggesting 
%significant
depletion of iron onto dust grains. In the high-metallicity galaxies
more than 90\% of iron is locked onto dust grains. On the other hand,
no significant iron depletion is present in the lowest-metallicity
galaxies. We also find evidence that dust destruction occurs on a time scale
of several Myr.

5. No galaxy with log N/O $\la$ --1.6 was found, suggesting that
the chemical evolution of 
local low-metallicity emission-line galaxies is different from that of some
high-redshift DLAs which have considerably lower
log N/O ($\sim$ --2.3). These DLAs are considered to be
truly young galaxies, with nitrogen produced only by massive stars. If this
interpretation is correct, then our sample of $\sim$ 400
dwarf emission-line galaxies contains no extremely young galaxy but rather
galaxies of ages $\ga$ 100 -- 300 Myr, required for the enrichment in nitrogen
by intermediate-mass stars.

6. Our data indicate an apparent increase of N/O with decreasing EW(H$\beta$),
best seen among galaxies from our sample that have
intermediate metallicities. We interpret this as evidence for gradual
enrichment of the H {\sc ii} region in nitrogen by massive stars from the most
recent starburst.  The magnitude of the observed effect is consistent with
that expected from current theoretical models for massive stars, if
account is taken for the density of the ejecta being larger than that of
the  H {\sc ii} regions.

\begin{acknowledgements}
Y. I. I. acknowledges the support of the Observatoire de Paris, and
Y. I. I. and N. G. G. thank the hospitality of the Observatoire de Gen\`eve
where part of this work was carried out.
Y. I. I., G. M. and N. G. G. acknowledge the support of the
Swiss SCOPE 7UKPJ62178 grant.
T. X. T. and Y. I. I. acknowledge the partial financial support of NSF
grant AST 02-05785. The research described in this publication was made
possible in part by Award No. UP1-2551-KV-03 of the US Civilian Research
\& Development Foundation for the Independent States of the Former
Soviet Union (CRDF).
%and by grant No. M/85-2004 of the Ukrainian Ministry of Education and Science.
All the authors acknowledge the work of the Sloan Digital Sky
Survey (SDSS) team.
Funding for the SDSS has been provided by the
Alfred P. Sloan Foundation, the Participating Institutions, the National
Aeronautics and Space Administration, the National Science Foundation, the
U.S. Department of Energy, the Japanese Monbukagakusho, and the Max Planck
Society. The SDSS Web site is http://www.sdss.org/.
     The SDSS is managed by the Astrophysical Research Consortium (ARC) for
the Participating Institutions. The Participating Institutions are The
University of Chicago, Fermilab, the Institute for Advanced Study, the Japan
Participation Group, The Johns Hopkins University, the Korean Scientist Group,
Los Alamos National Laboratory, the Max-Planck-Institute for Astronomy (MPIA),
the Max-Planck-Institute for Astrophysics (MPA), New Mexico State University,
University of Pittsburgh, University of Portsmouth, Princeton University, the
United States Naval Observatory, and the University of Washington.
\end{acknowledgements}


\begin{thebibliography}{}

\bibitem[Abazajian et al. (2005)]{A05} Abazajian, K., et al. 2005, \aj,
129, 1755

\bibitem[Aller (1984)]{A84} Aller, L. H. 1984, Physics of Thermal Gaseous
Nebulae (Dordrecht: Reidel)

%\bibitem[1989]{AG89} Anders, E., \& Grevesse, N. 1989, 
Geochim.Cosmochim.Acta, 53, 197

\bibitem[Bai et al. (2004)]{B04} Bai, G. S., Zhao, G., Chen, Y. Q., et al.
2004, \aap, 425, 671

%\bibitem[1972]{B72} Brocklehurst, M. 1972, MNRAS, 157, 211

\bibitem[Carretta et al. (2000)]{C00} Carretta, E., Gratton, R. G., \&
Sneden, C. 2000, \aap, 356, 238

\bibitem[Centuri\'on et al. (2003)]{C03} Centuri\'on, M., Molaro, P.,
Vladilo, G., P\'eroux, C., Levshakov, S. A., \& D'Odorico, V. 2003,
\aap, 403, 55

%\bibitem[Chiappini et al. (2003)]{CMM03} Chiappini, C., Matteucci, F.,
%\& Meynet, G. 2003, \aap, 410, 257

%\bibitem[1988]{CF88} Cota, S. A., \& Ferland, G. J. 1988, ApJ, 326, 889

%\bibitem[2003]{D02} Depagne, E., Hill, V., Spite, M., Spite, F.,
%Plez, B.,  et al. 2002, A\&A, 390, 187

\bibitem[Dessauges-Zavadsky et al. (2004)]{D04} Dessauges-Zavadsky, M.,
Calura, F., Prochaska, J. X., D'Odorico, S., \& Matteucci, F. 2004, \aap, 416,
79

%\bibitem[1993]{E93} Edvardsson, B., Andersen, J., Gustafsson, B.,
%Lambert, D. L., Nissen, P. E., \& Tomkin, J. 1993, A\&A, 275, 101

\bibitem[Fillipenko (1982)]{F82} Fillipenko, A. V. 1982, \pasp, 94, 715

%\bibitem[Fricke et al. (2001)]{F01} Fricke, K. J., Izotov, Y. I.,
%Papaderos, P., Guseva, N. G., \& Thuan, T. X. 2001, \aj, 121, 66

\bibitem[Garnett (1990)]{G90} Garnett, D. R. 1990, \apj, 363, 142

\bibitem[Garnett (1992)]{G92} Garnett, D. R. 1992, \aj, 103, 1330

%\bibitem[Grevesse \& Noels (1996)]{GN96} Grevesse, N., \& Noels, A. 1996,
%in Cosmic Abundances, ASP Conference Series, vol. 99, eds. S. S. Holt and
%G. Sonneborn

%\bibitem[Guseva et al. (2000)]{G00} Guseva, N. G., Izotov, Y. I.,
%\& Thuan, T. X. 2000, \apj, 531, 776

%\bibitem[Guseva et al. (2001)]{G01} Guseva, N. G., Izotov, Y. I.,
%Papaderos, P., et al. 2001, \aap, 378, 756

%\bibitem[Guseva et al. (2003a)]{G03a} Guseva, N. G., Papaderos, P.,
%Izotov, Y. I., et al. 2003a, \aap, 407, 75

%\bibitem[Guseva et al. (2003b)]{G03b} Guseva, N. G., Papaderos, P.,
%Izotov, Y. I., et al. 2003b, \aap, 407, 91

\bibitem[Guseva et al. (2003)]{G03} Guseva, N. G., Papaderos, P.,
Izotov, Y. I., et al. 2003c, \aap, 407, 105

%\bibitem[2003]{HK03} Heckman, T. M., \& Kauffmann, G. 2003, in Star Formation
%through Time, Granada, Sep. 2002, ASP, Eds. E. Perez,
%R. Gonzalez Delgado and G. Tenorio Tagle, in press

\bibitem[Henry et al. (2000)]{H00} Henry, R. B. C., Edmunds, M. G.,
\& K\"oppen, J. 2000, \apj, 541, 660

%\bibitem[2000]{Ho00} Hopp, U., Engels, D., Green, R. F., et al. 2000, A\&AS,
%142, 417

%\bibitem[Hunt et al. (2003)]{H03} Hunt, L. K., Thuan, T. X.,
%\& Izotov, Y. I. 2003, \apj, 588, 281

%\bibitem[Izotov \& Thuan (1998a)]{IT98a} Izotov, Y. I., \& Thuan, T. X.
%1998a, \apj, 497, 227

%\bibitem[Izotov \& Thuan (1998b)]{IT98} Izotov, Y. I., \& Thuan, T. X.
%1998b, \apj, 500, 188

\bibitem[Izotov \& Thuan (1999)]{IT99} Izotov, Y. I., \& Thuan, T. X.
1999, \apj, 511, 639

\bibitem[Izotov \& Thuan (2004a)]{IT04a} Izotov, Y. I., \& Thuan, T. X.
2004a, \apj, 602, 200

\bibitem[Izotov \& Thuan (2004b)]{IT04b} Izotov, Y. I., \& Thuan, T. X.
2004b, \apj, 616, 768

\bibitem[Izotov et al. (1994)]{ITL94} Izotov, Y. I., Thuan, T. X.,
\& Lipovetsky, V. A. 1994, \apj, 435, 647

%\bibitem[1996]{I96} Izotov, Y. I., Dyak, A. B., Chaffee, F. H., Foltz, C. B.,
%Kniazev, A. Y., \& Lipovetsky, V. A. 1996, ApJ, 458, 524

\bibitem[Izotov et al. (1997a)]{ITL97} Izotov, Y. I., Thuan, T. X.,
\& Lipovetsky, V. A. 1997a, \apjs, 108, 1

%\bibitem[Izotov et al. (1997b)]{I97} Izotov, Y. I., Lipovetsky, V. A.,
%Chaffee, F. H., Foltz, C. B., Guseva, N. G., \& Kniazev, A. Y.
%1997, \apj, 476, 698

%\bibitem[Izotov et al. (1999)]{I99} Izotov, Y. I., Chaffee, F. H.,
%Foltz, C. B., et al. 1999, \apj, 527, 757

%\bibitem[Izotov et al. (2001a)]{I01a} Izotov, Y. I., Chaffee, F. H.,
%\& Green, R. F. 2001a, \apj, 562, 727

%\bibitem[Izotov et al. (2001b)]{I01b} Izotov, Y. I., Chaffee, F. H.,
%\& Schaerer, D. 2001b, \aap, 378, L45

\bibitem[Izotov et al. (2004)]{I04} Izotov, Y. I., Stasi\'nska, G.,
Guseva, N. G., \& Thuan, T. X. 2004, \aap, 415, 87

\bibitem[Kauffmann et al. (2003)]{K03} Kauffmann, G., Heckman, T. M.,
Tremonti, C., et al. 2003, \mnras, 346, 1055

\bibitem[Kniazev et al. (2003)]{K04a} Kniazev, A. Y., Grebel, E. K.,
Hao, L., Strauss, M. A., Brinkmann, J., \& Fukugita, M. 2003, \apj, 593, 73

\bibitem[Kniazev et al. (2004)]{K04b} Kniazev, A. Y., Pustilnik, S. A.,
Grebel, E. K., Lee, H., \& Pramskij, A. G. 2004, \apjs, 153, 429

\bibitem[Leitherer et al. (1999)]{L99} Leitherer, C., Schaerer, D.,
Goldader, J. D., Gonzalez Delgado, R. M., Robert, C., Kune D. F.,
de Mello, D. F., Devost, D., \& Heckman, T. M. 1999, \apjs, 123, 3

%\bibitem[1999]{Li99} Lipovetsky, V. A., Chaffee, F. H., Izotov, Y. I., et al.
%1999, ApJ, 519, 177

\bibitem[Lodders (2003)]{L03} Lodders, K. 2003, \apj, 591, 1220

%\bibitem[Lu et al. (1998)]{L98} Lu, L., Sargent, W. L. W., \& Barlow, T. A.
%1998, \aj, 115

%\bibitem[Maeder (1992)]{M92} Maeder, A. 1992, \aap, 264, 105

\bibitem[Meynet \& Maeder (2005)]{MM05} Meynet, G., \& Maeder, A. 2005,
\aap, 429, 613

\bibitem[Molaro (2005)]{M05} Molaro, P. 2005, preprint astro-ph/0503214

\bibitem[Oke (1990)]{O90} Oke, J. B. 1990, \aj, 99, 1621

\bibitem[Pilyugin (1993)]{P92} Pilyugin, L. S. 1992, \aap, 260, 58

\bibitem[Ramsbottom et al. (1996)]{R96} Ramsbottom, C. A., Bell, K. L., \&
Stafford, R. P. 1996, ADNDT, 63, 57

%\bibitem[1989]{O89} Osterbrock, D. E. Astrophysics of gaseous nebulae and
%active galactic nuclei (University Science Books: Mill Valley)

\bibitem[Rodriguez \& Rubin (2005)]{R05} Rodriguez, M., \& Rubin, R. H. 2005,
\apj, 626, 900

%\bibitem[Sargent \& Searle (1970)]{SS70} Sargent, W. L. W., \& Searle, L.
%1970, \apj, 162, L155

%\bibitem[2003]{S03} Skillman, E., C\^ot\'e, S., \& Miller, B. W. 2003,
%AJ, 125, 610

\bibitem[Smith et al. (2002)]{S02} Smith, L. J., Norris, R. P. F.,
\& Crowther, P. A., 2002, \mnras, 337, 1309

\bibitem[Spite et al. (2005)]{Sp05} Spite, M., Cayrel, R., Plez, B., et al.
2005, \aap, 430, 655

\bibitem[Stasi\'nska (1990)]{S90} Stasi\'nska, G. 1990, \aaps, 83, 501

\bibitem[Stasi\'nska (2005)]{S05} Stasi\'nska, G. 2005, \aap, 434, 507


\bibitem[Stasi\'nska \& Izotov (2003)]{SI03} Stasi\'nska, G., \&
Izotov, Y. I. 2003, \aap, 397, 71

%\bibitem[2002]{S02} Stoughton, C., Lupton, R. H., Bernardi, M., et al. 2002,
%AJ, 123, 485

\bibitem[Takeda (2003)]{T03} Takeda, Y. 2003, \aap, 402, 343

\bibitem[Thuan et al. (1995)]{TIL95} Thuan, T. X., Izotov, Y. I.,
\& Lipovetsky, V. A. 1995, \apj, 445, 108

%\bibitem[Thuan et al. (1999)]{TIF99} Thuan, T. X., Izotov, Y. I.,
%\& Foltz, C. B. 1999, \apj, 525, 105

\bibitem[York et al. (2000)]{Y00} York, D. G., Adelman, J.,
Anderson, J. E., Jr., et al. 2000, \aj, 120, 1579

%\bibitem[2000]{V00} van Zee, L. 2000, ApJ, 543, L31

\bibitem[Whitford (1958)]{W58} Whitford, A. E. 1958, \aj, 63, 201

\end{thebibliography}
\end{document}